\documentclass[fleqn,usenatbib]{mnras}

\usepackage{newtxtext,newtxmath}

\usepackage[T1]{fontenc}

\DeclareRobustCommand{\VAN}[3]{#2}
\let\VANthebibliography\thebibliography
\def\thebibliography{\DeclareRobustCommand{\VAN}[3]{##3}\VANthebibliography}

\usepackage{graphicx}	
\usepackage{amsmath}	
\usepackage{epstopdf}
\usepackage{color}
\usepackage{soul}

\newcommand{\vect}[1]{\boldsymbol{#1}}

\DeclareMathOperator{\sgn}{sgn}

\defcitealias{Cheng2018}{C18}
\defcitealias{Liu2018}{L18}
\defcitealias{ONeill2021}{O21}

\title[GMC Collisions. VIII. The Core Mass Function]{GMC Collisions As Triggers of Star Formation. VIII.\\The Core Mass Function}

\author[Hsu et al.]{
Chia-Jung Hsu$^{1}$\thanks{E-mail: chiajung@chalmers.se},
Jonathan C. Tan$^{1,2}$,
Duncan Christie$^{2,3}$,
Yu Cheng$^{2,4}$,
and Theo J. O'Neill$^{2,5}$
\\
$^{1}$Department of Space, Earth \& Environment, Chalmers University of Technology, Gothenburg, Sweden \\
$^{2}$Dept. of Astronomy, University of Virginia, Charlottesville, Virginia 22904, USA\\
$^{3}$Max Planck Inst. for Astronomy, Heidelberg, Germany\\
$^{4}$National Astronomy Observatory of Japan, Mitaka, Tokyo, Japan\\
$^{5}$Harvard-Smithsonian Center for Astrophysics, Cambridge, MA 02138, USA
}

\date{Accepted XXX. Received YYY; in original form ZZZ}

\pubyear{2023}

\begin{document}
\label{firstpage}
\pagerange{\pageref{firstpage}--\pageref{lastpage}}
\maketitle

\begin{abstract}
Compression in giant molecular cloud (GMC) collisions is a promising mechanism to trigger formation of massive star clusters and OB associations. We simulate colliding and non-colliding magnetised GMCs and examine the properties of prestellar cores, selected from projected mass surface density maps, including after synthetic {\it ALMA} observations. We then examine core properties, including mass, size, density, velocity, velocity dispersion, temperature and magnetic field strength. After four Myr, $\sim1,000$ cores have formed in the GMC collision and the high-mass end of the core mass function (CMF) can be fit by a power law $dN/d{\rm{log}}M\propto{M}^{-\alpha}$ with $\alpha\simeq0.7$, i.e., relatively top-heavy compared to a Salpeter mass function. Depending on how cores are identified, a break in the power law can appear around a few $\times10\:M_\odot$. The non-colliding GMCs form fewer cores with a CMF with $\alpha\simeq0.8$ to 1.2, i.e., closer to the Salpeter index. We compare the properties of these CMFs to those of several observed samples of cores. Considering other properties, cores formed from colliding clouds are typically warmer, have more disturbed internal kinematics and are more likely to be gravitational unbound, than cores formed from non-colliding GMCs. The dynamical state of the protocluster of cores formed in the GMC-GMC collision is intrinsically subvirial, but can appear to be supervirial if the total mass measurement is affected by observations that miss mass on large scales or at low densities.
\end{abstract}

\begin{keywords}
hydrodynamics -- methods:numerical -- ISM:clouds -- stars:formation
\end{keywords}



\section{Introduction}

Collisions between giant molecular clouds (GMCs) have been proposed as a mechanism for triggering star formation, especially cluster formation and massive star formation \citep[e.g.,][]{Scoville1986}, with the collision quickly assembling large amounts of gas in a compact region.  Simulations of galactic disks have found that these events can occur on timescales much shorter ($\sim10-20\%$) than the local orbital period \citep{Tasker2009, Dobbs2015, Li2018}, and it has been proposed that collisions could explain the relationship between the star formation rate and the gas mass surface density divided by orbital time, i.e., the ``Dynamical Kennicutt-Schmidt'' relation \citep{Tan2000, Tan2010, Tasker2009, Suwannajak2014}.
Observations of molecular gas around some local massive young stellar clusters, in particular via analysis of CO channel maps, have identified a number of candidates for cloud-cloud collisions \citep[e.g.,][]{Furukawa2009, Fukui2014, 2017arXiv171101695F, Bisbas2018}.

In this paper, the eighth in a series investigating numerical models of magnetized GMC collisions, we present our highest resolution simulations to date, which enable us to identify structures that may be comparable with pre-stellar cores (PSCs). These have been defined theoretically to be self-gravitating gas structures that collapse to a central disk that forms a single or small $N$ multiple by disk fragmentation \citep[e.g.,][]{Tan2014}. This core mass function (CMF) may have a direct connection to the stellar initial mass function (IMF) \citep[e.g., see review by][]{Offner2014}.

Our goal in this paper is to extract CMFs from our simulation outputs with methods that closely follow those that are used in observational studies. We will then compare our simulated CMFs with those derived in observed regions \citep[e.g.,][]{Cheng2018, Liu2018,ONeill2021} to examine how well or how poorly they agree. We also examine how the CMFs depend on certain simulation and analysis properties, especially: colliding versus non-colliding GMCs; evolutionary stage; choice of dendrogram parameters; use of a density threshold to define material belonging to a core; and {\it ALMA}-like spatial filtering.

We provide a brief summary of the initial conditions and numerical methods in \S\ref{sec:method}, then present our results in \S\ref{sec:result}. We summarize in \S\ref{sec:discussion}.


\section{Numerical Simulations}
\label{sec:method}

The simulations presented here are based on the model presented in \citet[][hereafter Paper II]{Wu2017}, with updates to the heating and cooling functions discussed in \citet[][hereafter Paper IV]{Christie2017}. Here, we outline the main features of this setup, but refer the reader to these papers for more detailed descriptions. Using the \texttt{Enzo} magnetohydrodynamics code \citep{Wang2008, Wang2009, Brummel-Smith2019} with a simulation domain of $(128\,{\rm pc})^3$, two molecular clouds of radius $R_{\rm GMC} = 20\,{\rm pc}$ are initialized with an impact parameter of $b=0.5R_{\rm GMC}$. The clouds start with an uniform particle number density $n_{\rm particle} = 50\,{\rm cm^{-3}}$, which, given an adopted value of $n_{\rm He}=0.1 n_{\rm H}$, corresponds to a H nuclei number density of $n_{\rm H} = 83.3\,{\rm cm^{-3}}$ (previous papers in this series mistakenly listed $n_{\rm H}=100\:{\rm cm}^{-3}$ for this value). 
The density of ambient gas is set to be ten times smaller than that of the GMCs, i.e., $n_{\rm H} = 8.3\,{\rm cm^{-3}}$.
The clouds are embedded in a background FUV radiation field equivalent to four Habings, i.e., $G_0=4$, which is attenuated by an approximate local density-$A_V$ relation. The cosmic ray ionization rate yields a primary ionization rate of $10^{-16}\:{\rm s}^{-1}$, applied uniformly through the domain.
An uniform magnetic field with strength $B_0 = 10\,{\rm \mu G}$ is initialized with an orientation of $60^\circ$ relative to the collision axis. These initial conditions were considered the fiducial colliding and non-colliding cases in the previous papers in this series, although stronger field cases have also been considered by \citet{Wu2020}.

As in previous papers, we consider both colliding and non-colliding cases. In both, each cloud is initialized with a turbulent velocity field with a three-dimensional power spectrum following the relation $v_k^2 \propto k^{-4}$ and an initial sonic Mach number for the turbulence of $M_s = 23$ (assuming $T=15\,{\rm K}$). An identical velocity field to that used in Papers II and IV is adopted. To initialize the colliding case, the gas is given an additional velocity contribution of $-\frac{1}{2}\sgn(x)v_{\rm rel}\vect{\hat{x}}$ where $v_{\rm rel}=10\,{\rm km\, s^{-1}}$ is the relative velocity between the clouds. While no additional velocity contribution is included in the non-colliding case, some small relative motions between the clouds do develop due to mutual gravitational attraction, but these do not lead to collision during the time frame of the simulations. 

The base resolution has been increased to $256^3$ (from $128^3$ in Papers II and IV) with 5 levels of refinement using the requirement that the Jeans length be resolved by 8 zones, resulting in a grid size of $0.015625\,{\rm pc}$, i.e., 3,200 au, for the most refined grid. This is a factor of four increase in linear resolution over Paper IV and a factor of eight increase over Paper II. As in Papers II and IV, the simulations are run to 4~Myr.

We note some caveats and limitations of the simulations here. The Jeans length is not resolved for very dense structures within the collapse. At a temperature of $T=10\,{\rm K}$, the Jeans length $\lambda_J = (\pi c_s^2 / [G\rho] )^{1/2} = (\pi \gamma k T / [G \rho \mu m_{\rm H}])^{1/2} $ is no longer resolved by 8 zones at a density of $n_{\rm H} \simeq 6.69 \times 10^4\,{\rm cm^{-3}}$ if adopting $\gamma = 7/5$. However, we note that since the simulations also include magnetic field support, the Jeans length should be replaced by the 
magneto-Jeans fragmentation scale, $\lambda_{\rm MJ} = [\pi(v_{\rm A}^2\cos^2\theta + c_{\rm s}^2)/G\rho]^{1/2}$, which depends on the orientation angle $\theta$ of the perturbation relative to the magnetic field. For perturbations oriented along the magnetic field, the fragmentation scale remains the Jeans scale. For perturbations perpendicular to the magnetic field, and assuming the magnetic field strength scales as $B = 10 (n_{\rm H}/83\,{\rm cm^{-3}})^{1/2}\,{\rm \mu G}$ (with density here normalized to the initial $n_{\rm H}$ in the clouds),
the magneto-Jeans lengthscale is resolved by 8 zones up to densities $n_{\rm H} \simeq 5.59 \times 10^6\,{\rm cm^{-3}}$. This suppression of fragmentation limits the amount of artificial fragmentation associated with under-resolving the Jeans scale. 

In addition, the simulations do not include the formation of stars (via star particle creation; see \citealt{Wu2017a, Wu2020} [Paper III, VII]), which is a deliberate choice to focus on PSCs and since the star formation process cannot be resolved and would need to rely on an uncertain sub-grid model. Thus, it is possible that cores build up to masses and densities at which, in reality, a star would already have formed inside. We will check this after the fact by comparing the core densities achieved with those in known PSCs. Furthermore, without star formation there is also no protostellar feedback, especially MHD outflows and radiative heating, both on the scale of the core and the surrounding clump. We discuss later how these simulation caveats should be considered in the interpretation of the results.


\subsection{Core Identification Method}

Our goal is to identify cores from the projected mass surface density map.
To facilitate comparison with observational results, we follow the methods of \citet{Cheng2018} \citep[see also][]{Liu2018, ONeill2021} for identifying cores from 2D images, based on the dendrogram algorithm implemented in the {\em astrodendro} package \citep{Rosolowsky2008}. In our case, we analyze images of the mass surface density, $\Sigma$, of the simulated structures. In the above observational studies, the analysis was done on $\sim$1.3~mm continuum images. 
Assuming that the continuum emission is due to optically thin thermal emission from dust having uniform temperature and emissivity properties, then the 1.3~mm continuum map has direct correspondence with our mass surface density map. Other types of observational studies estimate mass surface density via dust extinction \citep[e.g.,][]{2012ApJ...754....5B}, which is independent of local temperature, although such maps have not yet been utilizaed to estimate the CMF.

The dendrogram algorithm requires three main parameters: minimum mass surface density, minimum increment of mass surface density, and minimum area. In the observational study of \citet{Cheng2018}, the level of the continuum noise in their image that has $\sim1\arcsec$ resolution (i.e., a FWHM beam diameter of about 1\arcsec) was 0.45~mJy/beam. In the study of \citet{Liu2018}, the equivalent noise level for a similar angular resolution image was about 0.2~mJy/beam. In \citet{ONeill2021}, the noise levels range from 0.13 to 1.38~mJy/beam, but most of them are lower than 0.5~mJy/beam. For the fiducial assumptions of conversion of mm continuum flux into mass surface density (i.e., dust temperature of 20~K; gas to refractory component dust mass ratio of 141 \citep{2011piim.book.....D}; opacity per unit dust mass of $\kappa_{\rm 1.3mm,d}=0.899\:{\rm cm^2\:g^{-1}}$ \citep{1994A&A...291..943O}), 0.45~mJy/beam corresponds to $\Sigma=0.122\:{\rm g\:cm}^{-2}$. The threshold for identifying cores in the observational studies was $4\sigma$, i.e., $0.49\:{\rm g\:cm}^{-2}$, with an increment of $1\sigma$. The equivalent thresholds in the study of \citet{Liu2018} are about a factor of two lower. Thus, when we examine the simulated core populations, we will also explore the effect on the CMF from varying $\Sigma_{\rm min}$ from our fiducial value of 0.1$\:{\rm g\:cm}^{-2}$ to 0.05$\:{\rm g\:cm}^{-2}$, thus spanning the range of these observational studies. In the fiducial case, we adopt a minimum mass surface density threshold $\Sigma_{\rm min} = 0.1\,{\rm g\, cm^{-2}}$. We then search for fragmentation in increments of mass surface density $\delta_{\rm min} = 0.025\,{\rm g\, cm^{-2}}$. 

For the third parameter, we require that the minimum projected area, $A_{\rm min}$, of each core is at least two zones at the finest grid scale, i.e., an area of $2.44\times 10^{-4}\,\,{\rm pc^2}$. This choice of two contiguous pixels is equivalent to $3.32$ square arcseconds when the source is at 2.5~kpc (as in the case of G286 studied by \citealt{Cheng2018}) or 0.83 square arcseconds for sources at 5~kpc, typical of the most distant IRDCs in the sample of \citet{Liu2018}. However, these observational studies have employed a minimum core angular area of 0.5 beam areas in their fiducial cases, i.e., about 0.8 square arcseconds, which, especially in the case of the more nearby G286 source, is smaller than we can achieve with the simulations. Moreover, these observational studies can detect down to this level of minimum area with no shape dependence, while the minimum areas of the cores detected in the simulated images are pixelated to have an axis ratio of 2:1.

Given that the focus of the paper is on cores identified in
projection, we consider mass-weighted quantities, i.e.,
\begin{equation}
\label{Eqn:MassWeight}
\langle X\rangle_M= \frac{1}{M}\int dA \int X \rho ds,
\end{equation}
where $A$ is the area of the core in projection, $s$ is the normal along the line of sight, and $M$ is the total mass of the considered structure. 
We will also consider cases where material needs to be above a threshold density to be counted as part of a core structure.

\subsection{ALMA Synthetic Observation}
\label{sec:alma}


To make more direct comparison of our simulation results with observational studies, we generate synthetic ALMA observations, which we refer to as ``ALMA filtered'', and then perform our dendrogram core-finding procedure on these images. To generate the synthetic observations, we produce a flux map at 1.3~mm assuming optically thin thermal dust emission derived from mass surface density maps. For simplicity we assume a temperature of 20~K, which is expected to be representative of the average temperature in protostellar cores \citep[e.g.][]{Zhang2015} (note, our simulations do not include protostellar heating). An opacity per unit dust mass $\kappa_{1.3\, {\rm mm}} = 0.899\, {\rm cm^2/g}$ \citep{1994A&A...291..943O} is adopted, along with a gas-to-refractory-component-dust ratio of 141 \citep{2011piim.book.....D}. These assumptions for temperature, opacity and dust-to-gas mass ratio are the same as those made in the observational studies of \citet{Cheng2018}, \citet{Liu2018} and \citet{ONeill2021} for converting observed 1.3~mm flux into mass surface density. We assume a distance of 5~kpc and adopt an ALMA Band 6 compact configuration with an angular resolution of $\sim$1.5\arcsec, which corresponds to a linear resolution of about 7500~AU, i.e., roughly twice as large as the spatial resolution of the finest grid in our simulations.

Then the ALMA filtered images were produced using the CASA software \citep{2007ASPC..376..127M}: first, synthetic visibilities were created with the task {\it simobserve}. To compare with realistic observations, we assume the same phase center as the observational setup for the protocluster G286 \citep{Cheng2018}, which is located at (R.A.=10:38:33, decl.=-58:19:22). We perform a 3.54~s integration for each pointing with a 2~GHz bandwidth. The integration time helps us control the noise close to 0.025~$\rm g\:cm^{-2}$, i.e., the fiducial 1$\sigma$ increment used in dendrogram. The generated visibilities were then imaged and cleaned with the task {\it simanalyze}. To reduce the computational complexity, we have selectively sampled multiple regions of $768 \times 768$ pixels inside the map, which is equivalent to $\sim$144~$\rm pc^2$. To avoid missing dense cores in this region, we have included all subregions with pixels above a threshold of 0.1~$\rm g cm^{-2}$, which is a required condition for our fiducial core identification. These subregions overlap with each other by at least 64 cells on each boundary to reduce to error of edges. In each sample region, the task have a maximum number iteration of 1,000,000 and an upper threshold of cleaning of 0.84~mJy. The final output of the tasks reported the major and the minor beam sizes are $\sim 1.89$ and $1.54$ arcsec. The outputs without the primary beam correction are used in the analysis as it has a flat noise profile.

\section{Results}
\label{sec:result}

\subsection{Global Evolution}

\begin{figure*}
    \centering
    \includegraphics[width=\linewidth]{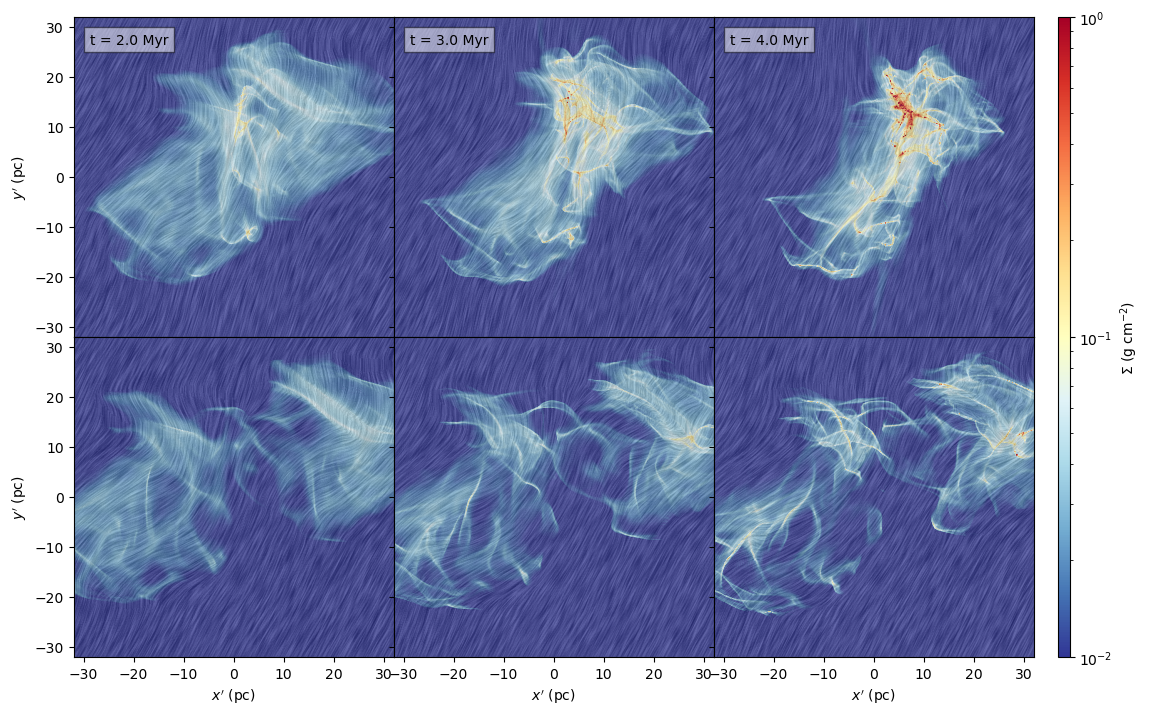}
    \caption{Global evolution of cloud structures. Here the mass surface density, as viewed along the $z'$ axis, is shown for the colliding (top) and non-colliding (bottom) cases at 2, 3, and 4~Myr (left to right). The mass-weighted magnetic field orientation is overlaid with the texture from line-integral-convolution method. 
    }
    \label{fig:surfdens}
\end{figure*}

Figure~\ref{fig:surfdens} shows the time evolution of both the colliding and non-colliding cases, with snapshots of mass surface density, $\Sigma$, shown at 2, 3 and 4~Myr. As in previous papers, visualization and analysis are done in a coordinate frame ($x', y', z'$) that is rotated by $15^\circ$ in each of the $\theta$ and $\phi$ directions from the collision axis, which minimizes the morphology of a compressed thin sheet formed from the collision of the uniform ambient medium.

The colliding case forms dense gas structures, including ``cores'' (see below), at relatively early times, driven initially by compression at the collision interface between the two clouds. At first, this resembles simulations of colliding flows \citep[e.g.,][]{Chen2018} where core formation proceeds in a thin sheet. However, as the simulation progresses, more density substructure develops and the collision becomes qualitatively less like simple colliding flows. Especially, the dense gas concentrated by the collision becomes self-gravitating, further concentrating the material. On the other hand, the non-colliding case takes longer to develop dense structures, with the initial turbulent velocity field being the main cause of generating density enhancements and significant amounts of dense structures not appearing until towards the end of the simulation.

\begin{figure*}
    \centering
    \includegraphics[width=\linewidth]{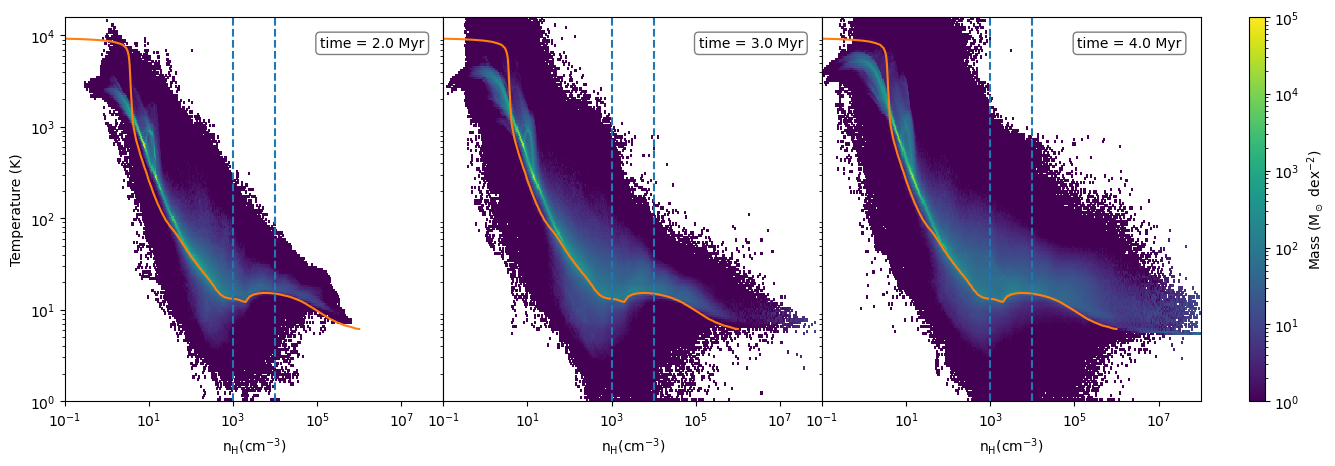}
    \includegraphics[width=\linewidth]{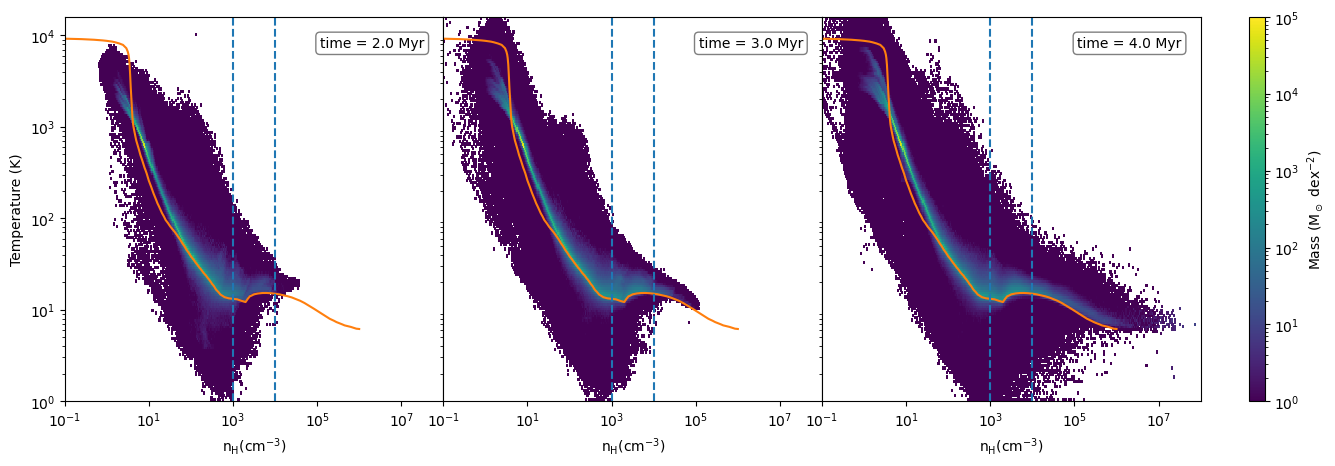}
    \caption{Two-dimensional histogram distribution of the density and temperature. The color indicate the mass distributed in that range. Panels are the results from 2, 3, and 4~Myr from left to right. (a) Top: colliding case (b) Bottom: non-colliding case.}
    \label{fig:hist2d-dens-temp-cc}
\end{figure*}

Figure~\ref{fig:hist2d-dens-temp-cc} shows the temperature-density phase diagram of the colliding and non-colliding simulations at 2, 3 and 4~Myr. We see that most mass is concentrated at conditions close to those expected from thermal equilibrium, given our implemented photo-dissociation region (PDR) and molecular cloud heating and cooling functions. However, we note a greater dispersion in temperatures at a given density in the GMC collision simulation compared to the non-colliding case, especially towards warmer temperatures. We attribute this to a greater degree of compressional heating in the colliding case, both from the GMC-GMC collision itself on large scales, but also from the more rapid accumulation of gas in localized dense cores. We will return to this point when discussing the thermal properties of dense cores identified in the simulations in \S\ref{sec:temp}.

\begin{figure*}
    \centering
    \includegraphics[width=\linewidth]{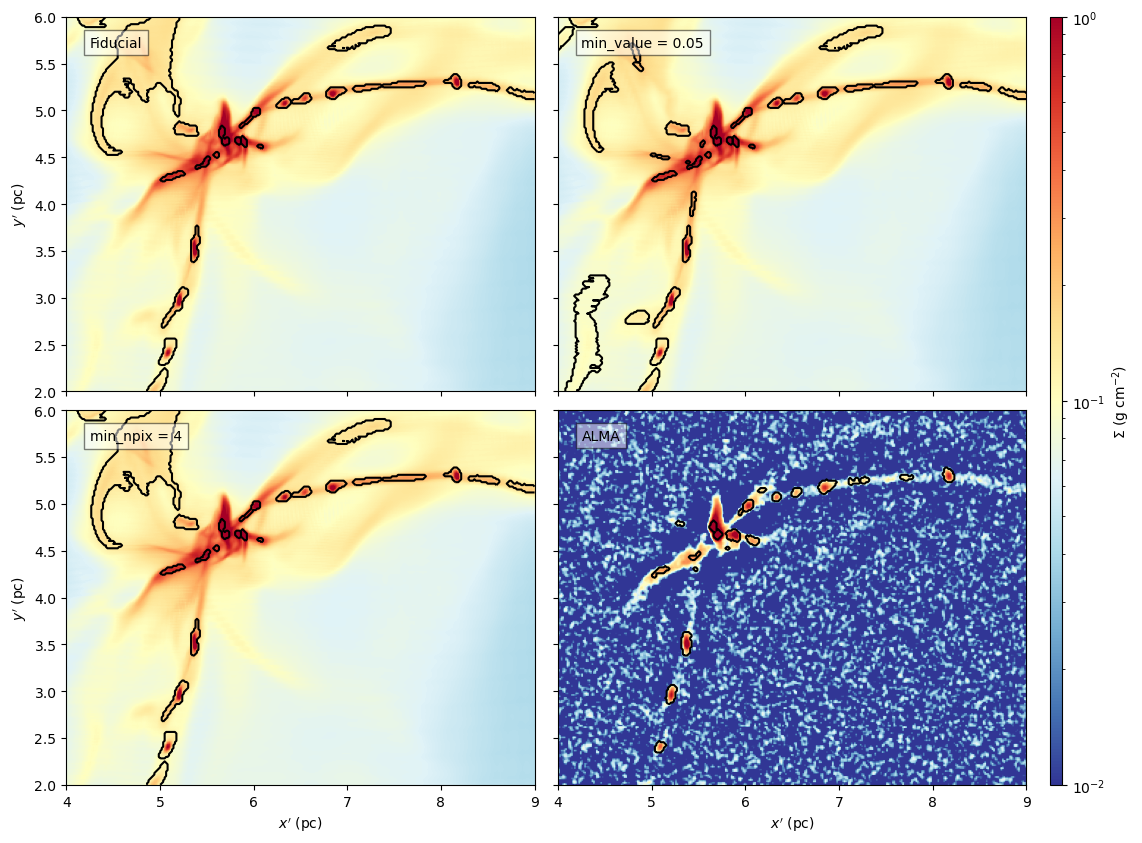}
    \caption{A zoomed-in view of the mass surface density of a dense region produced by the GMC-GMC collision. Cores identified by dendrogram are outlined with black contours. {\it (a) Top Left}: Dendrogram parameters for core identification are set to $\Sigma_{\rm min} = 0.1\:{\rm g\:cm}^{-2}$, $\delta_{\rm min}=0.025\:{\rm g\:cm}^{-2}$ and $A_{\rm min}=2$ pixels. {\it (b) Top Right}: As (a), but with dendrogram parameters set to $\Sigma_{\rm min} = 0.05\:{\rm g\:cm}^{-2}$, $\delta_{\rm min}=0.0125\:{\rm g\:cm}^{-2}$ and $A_{\rm min}=2$ pixels. {\it (c) Bottom Left}: As (a), but with dendrogram parameters set to $\Sigma_{\rm min} = 0.1\:{\rm g\:cm}^{-2}$, $\delta_{\rm min}=0.025\:{\rm g\:cm}^{-2}$ and $A_{\rm min}=4$ pixels. {\it (d) Bottom Right}: As (a), including fiducial dendrogram paraemters, but with the image processes to yield an ALMA synthetic observation (see text).
    }
    \label{fig:dendrodemo}
\end{figure*}

\begin{figure*}
    \centering
    \includegraphics[width=\linewidth]{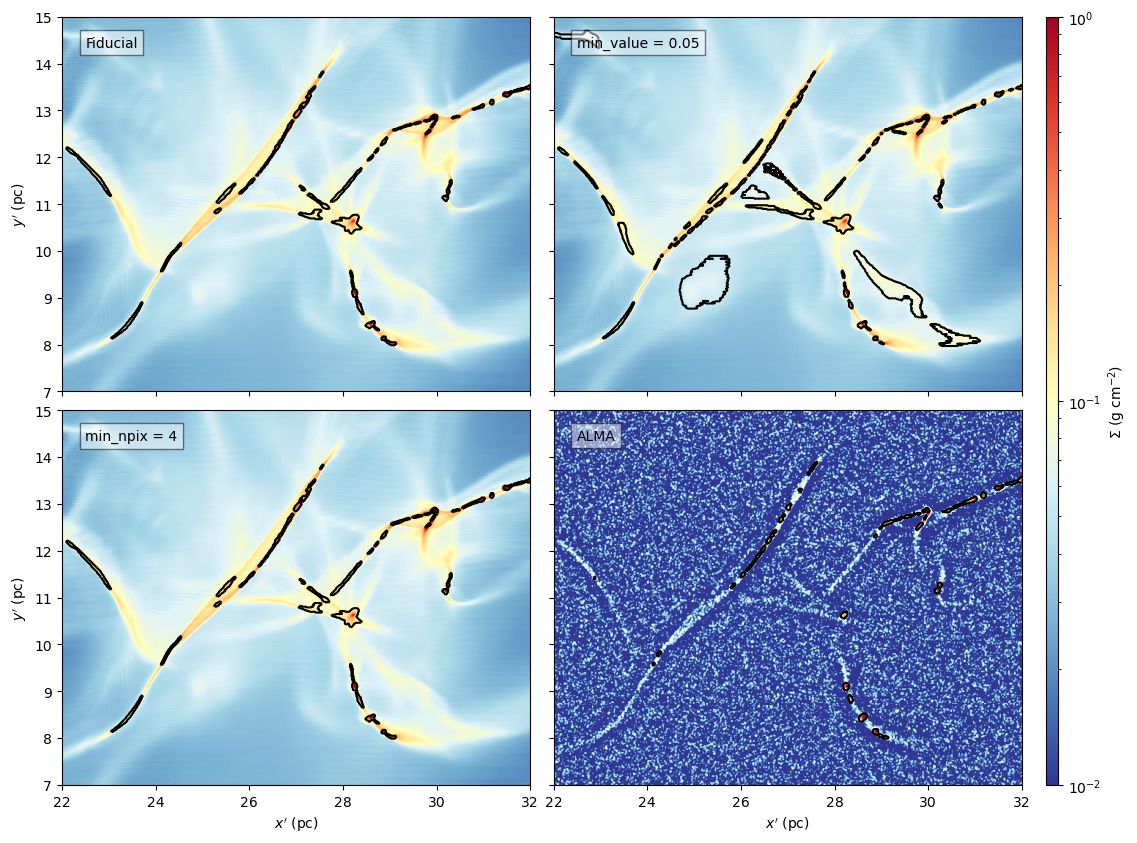}
    \caption{Same as Figure~\ref{fig:dendrodemo}, but now with a zoomed-in view of a dense region in the non-colliding case.}
    \label{fig:dendrodemo_nocol}
\end{figure*}



\subsection{Core Identification}
\label{sec:coreid}

Figure~\ref{fig:dendrodemo} shows a zoom-in of an example high density region formed in the colliding case at $t=4\,{\rm Myr}$. The cores identified by the dendrogram algorithm are outlined with black contours, including the effects of different choices of $\Sigma_{\rm min}$, $\delta_{\rm min}$ and $A_{\rm min}$. We note that a significant fraction of the cores are found along large-scale filamentary structures and that these cores are often themselves filamentary with a similar orientation. We see that lowering $\Sigma_{\rm min}$ leads, as expected, to identification of cores in lower $\Sigma$ regions and that these tend to be larger, more diffuse structures. The effect of doubling $A_{\rm min}$ from the fiducial case has very little effect on the number and type of core identified in this region. Figure~\ref{fig:dendrodemo} also shows the impact of ALMA filtering of this region, including on core identification. Now the effect is much more dramatic, primarily because large, extended structures are no longer present in the map. Consequences of this include that cores are smaller, less filamentary and confined to denser regions.

Figure \ref{fig:dendrodemo_nocol} shows the equivalent information as Figure~\ref{fig:dendrodemo}, but now for a region extracted from the non-colliding case. The same general trends for core identification are observed. We note that the density structures here include very thin, elongated filaments. Cores identified by dendrogram, especially in the case before ALMA filtering, can be extremely filamentary. 

\begin{figure}
    \centering
    \includegraphics[width=\columnwidth]{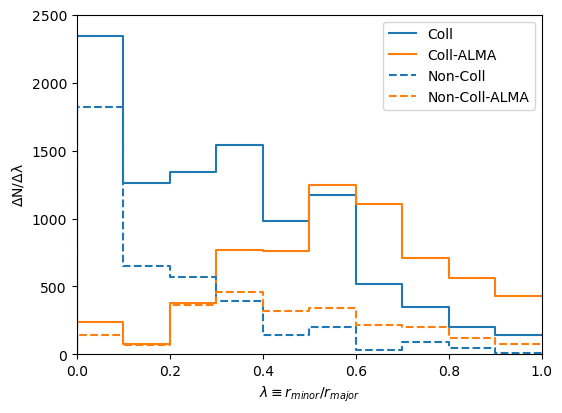}
    \caption{The distribution of the ratio, $\lambda$, between the minor axes ($r_{\rm minor}$) and major axes ($r_{\rm major}$) for cores identified in the colliding (solid lines) and non-colliding simulations (dashed lines) at $t=4$~Myr. Blue and orange lines show the results before and after ALMA filtering, respectively.
    }
    \label{fig:coreshape}
\end{figure}

To ascertain the degree to which the filtered observations suppress the elongated and filamentary cores within our sample, we examine the ratio of minor to major axes, $\lambda = r_{\rm minor}/r_{\rm major}$, with $r_{\rm minor}$ and $r_{\rm major}$ determined from the mass-surface-density-weighted second moments. The distributions of $\lambda$ for cores identified with and without ALMA filtering are 
shown in Figure~\ref{fig:coreshape}. For the collision simulation, the mean ratios are $0.314 \pm 0.245$ and $0.565 \pm 0.222$ for the original and ALMA-filtered cases, respectively.  For the non-colliding simulation, the mean ratios are $0.178 \pm 0.209$ and $0.460 \pm 0.229$ for the original and ALMA-filtered cases, respectively. Thus, we find that the distribution of $\lambda$ is strongly affected by ALMA filtering. Thus, when measuring this quantity observationally from interferometric data one should be aware of its potential dependence on the parameters of the observing set-up.




\subsection{Core Mass Function}
\label{sec:cmf}

\begin{figure*}
    \centering
    \includegraphics[width=\linewidth]{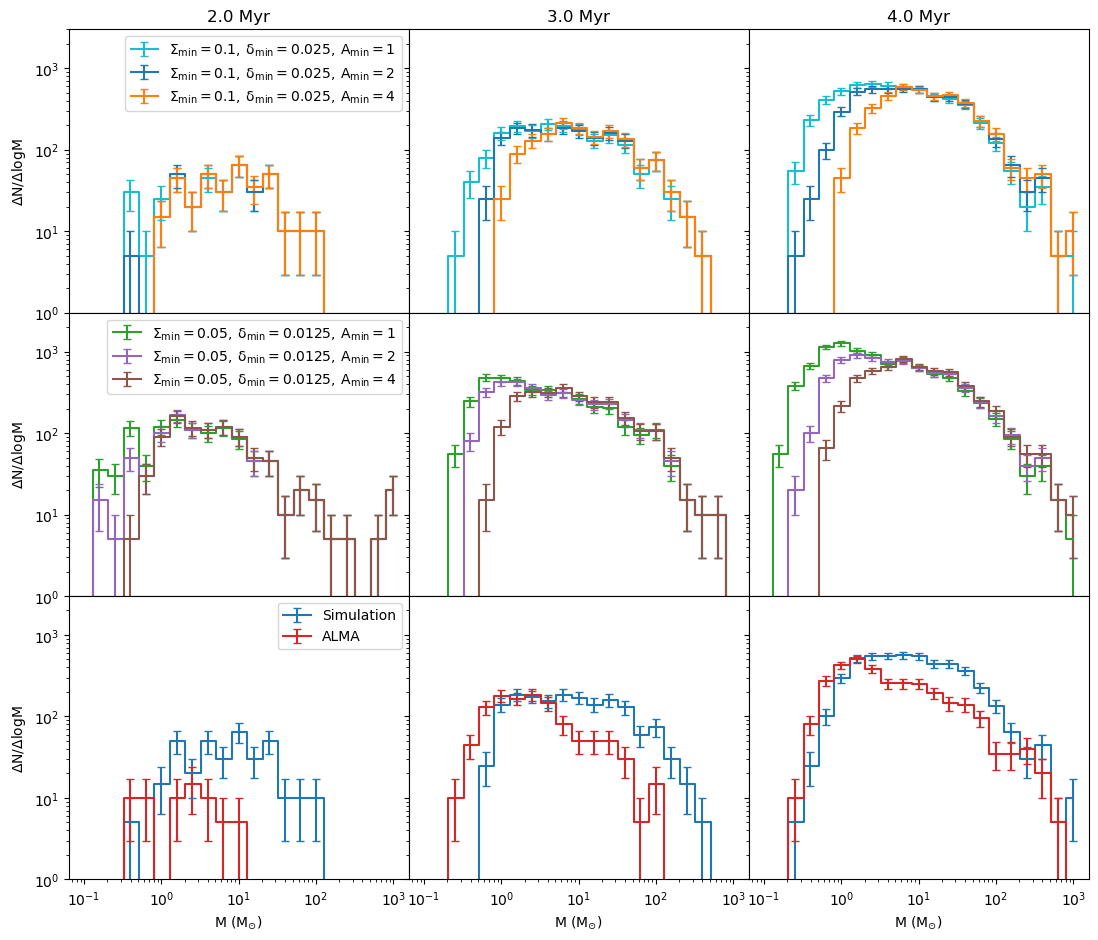}
    \caption{Time evolution of the core mass function (CMF) in the GMC-GMC collision simulation and its dependence on dendrogram parameters and ALMA filtering. {\it Left to Right:} CMFs at 2, 3 and 4~Myr. {\it First Row:} CMFs found using minimum mass surface density $\Sigma_{\rm min}=0.1\:{\rm g\:cm}^{-2}$ and minimum increment $\delta_{\rm min}=0.025\:{\rm g\:cm}^{-2}$. The minimum area varies from $A_{\rm min}=1$ (cyan), 2 (blue), to 4 (orange) pixels. The error bars indicate Poisson counting uncertainties. {\it Second Row:} As first row, but using $\Sigma_{\rm min}=0.05\:{\rm g\:cm}^{-2}$ and $\delta_{\rm min}=0.0125\:{\rm g\:cm}^{-2}$. The value of $A_{\rm min}$ varies from 1 (green), 2 (purple), to 4 (brown) pixels. {\it Third Row:} CMF of the original simulation data (blue) and after ALMA filtering (red).}
    \label{fig:dendrocmf}
\end{figure*}

In Figure~\ref{fig:dendrocmf}, we plot the CMFs found at 2, 3 and 4~Myr in the GMC collision simulation, exploring the effects of different dendrogram parameter choices and whether or not ALMA filtering has been applied to the projected image of the structures.
We have adopted a binning scheme identical to that of \citet{Cheng2018,Liu2018} and \citet{ONeill2021}, i.e., 5 bins per dex with bins centered on 1, 10, 100~$M_\odot$, etc. 
The blue line represents our fiducial case, with $\Sigma_{\rm min} = 0.1\:{\rm g\:cm}^{-3}$, $\delta_{\rm min} = 0.025\:{\rm g\: cm}^{-3}$ and $A_{\rm min} = 2$ pixels. 

As time evolves, the overall number of cores increases, i.e., with 69, 330, and 984 cores identified at 2, 3 and 4~Myr in the fiducial case. The maximum mass of the cores also increases. The high-mass end of the CMF appears to be approximately described by a power law distribution, which then exhibits a break at lower masses. This high-mass end of the CMF is relatively insensitive to the choice of dendrogram parameters. At lower masses, the CMF flattens further and then declines at masses below $\sim1\:M_\odot$. However, the precise location of the peak in the CMF depends on dendrogram parameters.
For example, comparing the first and second rows, we notice that a lower value of minimum density and increment causes the peak of the CMF to shift to smaller masses. 


Fiducial core identification in the ALMA filtered images yields 13, 228, and 629 cores at 2, 3 and 4~Myr, i.e., significantly smaller numbers than found in the original images.
In the third row, we see that ALMA postprocessing generally moves the peak of the CMF to smaller masses, i.e., close to $1 M_\odot$.
In the original CMFs, a break is apparent around a few $\times 10\:M_\odot$, but this feature is less clear after ALMA post-processing.

\begin{figure}
    \centering
    \includegraphics[width=\columnwidth]{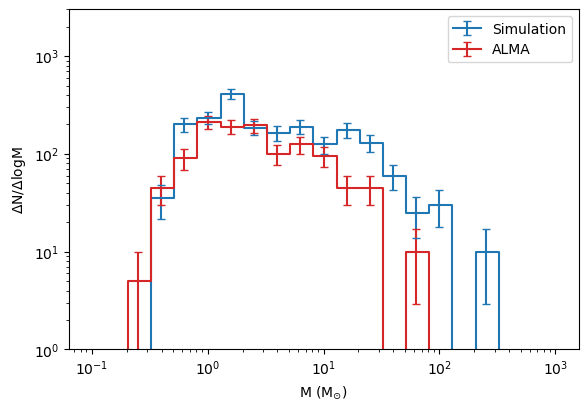}
    \caption{
    The core mass functions (CMFs) of the non-colliding GMCs simulation at 4~Myr. The CMF derived from the original simulation data is shown in blue. The CMF derived after ALMA filtering is shown in red. The error bars indicate Poisson counting uncertainties.
    }
    \label{fig:synthcmf_nocol}
\end{figure}

In the non-colliding simulation dense gas structures, including cores, take longer to form. With the fiducial method of core identification we find 0, 8 and 395 cores at 2, 3 and 4~Myr. Thus we focus on the CMF at 4~Myr in this simulation: Figure~\ref{fig:synthcmf_nocol} shows the original and ALMA-filtered CMFs of this case. The effect of ALMA filtering, where 231 cores are found, is similar to that seen in the colliding case, i.e., removing higher-mass cores and generally shifting the CMF to lower masses.


\begin{figure*}
    \centering
    \includegraphics[width=\linewidth]{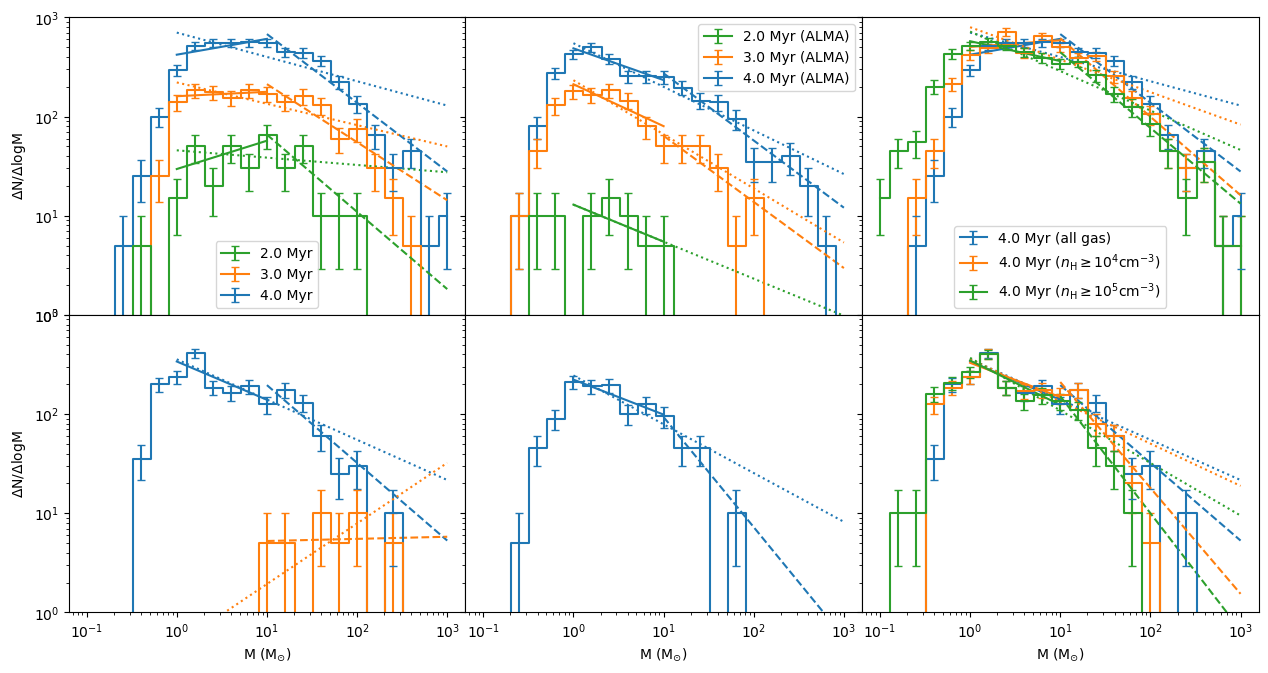}
    \caption{The core mass functions (CMFs) of colliding (top) and non-colliding (bottom) cases. {\it (a) Left column:} CMFs of original simulations, including at 2, 3 and 4~Myr when sufficient cores are present. {\it (b) Middle column:} CMFs of ALMA-filtered images. {\it (c) Right column} CMFs at 4~Myr under different density thresholds of $n_{\rm H}=10^4\:{\rm cm}^{-3}$ and $10^5\:{\rm cm}^{-3}$, as well as the case of no density threshold for reference. In each panel the error bars show Poisson counting uncertainties. The best power law fits over various mass ranges are also shown (see text and Table~\ref{tbl:cmf}). 
    }
    \label{fig:cmf-sim-alma-thres}
\end{figure*}

The left column of Figure~\ref{fig:cmf-sim-alma-thres} shows the time evolution of CMFs for both the colliding and non-colliding cases from 2 to 4~Myr, along with various power law fits of the form
\begin{equation}
\frac{d N}{d\log M} \propto  M^{-\alpha}.
\end{equation}
The fiducial \citet{1955ApJ...121..161S} initial mass function of stars has an index $\alpha=1.35$. To make direct comparison with the observational CMF results of \citet{Cheng2018}, \citet{Liu2018} and \citet{ONeill2021}, we fit the power law to the range $M / M_\odot \geq 1$, whose index we refer to as $\alpha_1$. In addition, as \citet{ONeill2021} claim that there is a break around $\sim10\:M_\odot$, we also examine the power law fits in the range $1 \leq M / M_\odot \leq 10$ (i.e., to derive index $\alpha_{1-10}$) and $M / M_\odot \geq 10$ (i.e., to derive index $\alpha_{10}$). The fitting procedure follows that of \citet{Cheng2018}, which fits the power law in logarithmic space, adopts Poisson errors, sets empty bins to 0.1 with errors of 1 dex, and sets bins with count of 1 to have an upper error of $\log 2$ dex and a lower error of 1 dex. We also make fits to ALMA-filtered images of the clouds. 

In addition, to ascertain to what degree the cores are affected by the presence of lower density gas along the line of sight, we consider cases where we recalculate core masses including only gas above a given density threshold. Note, here we still use the core contours identified using the full mass surface density image to make the comparison more direct on a core by core basis. However, we note that the density threshold condition can make some cores disappear. The results for these CMFs and their power law fits are shown in Figure~\ref{fig:cmf-sim-alma-thres} and listed in Table~\ref{tbl:cmf}.




\begin{table*}
    \centering
    \caption{Core Mass Function Properties}
    \label{tbl:cmf}
    \begin{tabular}{lccccccc}
    Case & \#Cores & \multicolumn{3}{c|}{Mass} & \multicolumn{3}{c|}{Power Law Indices, $\alpha$} \\
    \cline{3-5} \cline{6-8}
    & & $M_{\rm max}$ &  $\langle M\rangle_a$ & $\langle M\rangle_g$ & $\alpha_1 (M/\rm M_{\odot} \geq 1)$ & $\alpha_{1-10}(1 \leq M/\rm M_{\odot} \leq 10)$ & $\alpha_{10}(M/\rm M_{\odot} \geq 10)$ \\
    \hline
    Colliding Case & & & & & & \\
    \hline
    2~Myr & 69 & $98.5$ & $14.7$ & $7.33$ & $0.074 \pm 0.115$ & $-0.286 \pm 0.202$ & $0.779 \pm 0.289$ \\
    2~Myr ($M \geq 1\rm M_\odot$) & 67 & $98.5$ & $15.1$ & $7.88$ & & & \\
    2~Myr (ALMA) & 13 & $8.22$ & $2.68$ & $1.79$ & $0.373 \pm 0.492$ & $0.373 \pm 0.492$ & -- \\
    2~Myr (ALMA, $M \geq 1\rm M_\odot$) & 9 & $8.22$ & $3.62$ & $3.04$ & & & \\
    \hline
    3~Myr & 330 & $470$ & $23.0$ & $7.94$ & $0.215 \pm 0.039$ & $-0.044 \pm 0.092$ & $0.585 \pm 0.091$ \\
    3~Myr ($M \geq 1\rm M_\odot$) & 312 & $470$ & $24.3$ & $9.03$ & & & \\
    3~Myr (ALMA) & 228 & $123$ & $6.76$ & $2.58$ & $0.546 \pm 0.066$ & $0.423 \pm 0.117$ & $0.667 \pm 0.247$ \\
    3~Myr (ALMA, $M \geq 1\rm M_\odot$) & 168 & $123$ & $8.93$ & $4.21$ & & & \\
    \hline
    4~Myr & 984 & $947$ & $24.8$ & $7.91$ & $0.245 \pm 0.024$ & $-0.160 \pm 0.055$ & $0.693 \pm 0.051$ \\
    4~Myr ($M \geq 1\rm M_\odot$) & 936 & $947$ & $26.0$ & $8.94$ & & & \\
    *4~Myr ($\rm n_H \geq 10^4\,cm^{-3}$) & 983 & 811 & 19.6 & 5.96 & $0.327 \pm 0.024$ & $-0.066 \pm 0.053$ & $0.791 \pm 0.055$ \\
    *4~Myr ($\rm n_H \geq 10^4\,cm^{-3}, M \geq 1\rm M_\odot$) & 890 & 811 & 21.6 & 7.46 & & & \\
    *4~Myr ($\rm n_H \geq 10^5\,cm^{-3}$) & 933 & 808 & 16.7 & 3.66 & $0.398 \pm 0.026$ & $0.201 \pm 0.056$ & $0.763 \pm 0.061$ \\
    *4~Myr ($\rm n_H \geq 10^5\,cm^{-3}, M \geq 1\rm M_\odot$) & 720 & 808 & 21.5 & 6.61 & & & \\
    4~Myr (ALMA) & 629 & $628$ & $17.8$ & $4.02$ & $0.440 \pm 0.030$ & $0.320 \pm 0.065$ & $0.678 \pm 0.073$ \\
    4~Myr (ALMA, $M \geq 1\rm M_\odot$) & 513 & $628$ & $21.7$ & $6.02$ & & & \\
    \hline
    Non-Colliding Case & & & & & & \\
    \hline
    3~Myr & 8 & $283$ & $78.4$ & $50.8$ & $-0.611 \pm 0.317$ & -- & $-0.021 \pm 0.262$ \\
    3~Myr ($M \geq 1\rm M_\odot$) & 8 & $283$ & $78.4$ & $50.8$ & & & \\
    \hline
    4~Myr & 395 & $227$ & $9.95$ & $3.59$ & $0.406 \pm 0.044$ & $0.386 \pm 0.084$ & $0.784 \pm 0.144$ \\
    4~Myr ($M \geq 1\rm M_\odot$) & 327 & $227$ & $11.9$ & $5.08$ & & & \\
    *4~Myr ($\rm n_H \geq 10^4\,cm^{-3}$) & 395 & 123 & 7.02 & 2.99 & $0.426 \pm 0.049$ & $0.346 \pm 0.082$ & $1.068 \pm 0.179$ \\
    *4~Myr ($\rm n_H \geq 10^4\,cm^{-3}, M \geq 1\rm M_\odot$) & 314 & 123 & 8.66 & 4.46 & & & \\
    *4~Myr ($\rm n_H \geq 10^5\,cm^{-3}$) & 372 & 70.0 & 4.81 & 2.17 & $0.530 \pm 0.058$ & $0.448 \pm 0.084$ & $1.180 \pm 0.265$ \\
    *4~Myr ($\rm n_H \geq 10^5\,cm^{-3}, M \geq 1\rm M_\odot$) & 269 & 70.0 & 6.41 & 3.64 & & & \\
    4~Myr (ALMA) & 231 & $54.7$ & $4.80$ & $2.53$ & $0.492 \pm 0.074$ & $0.353 \pm 0.097$ & $1.111 \pm 0.352$ \\
    4~Myr (ALMA, $M \geq 1\rm M_\odot$) & 181 & $54.7$ & $5.93$ & $3.65$ & & & \\
    \hline
    Observational Comparisons & & & \\
    \hline
    \citet{Cheng2018} (Raw) & 76 & 80.2 & 2.79 & 1.10 & $1.108 \pm 0.197$ & $1.091 \pm 0.249$ & $1.084\pm 0.834$ \\
    \citet{Liu2018} (Raw) & 107 & 178 & 7.31 & 2.86 & $0.495 \pm 0.100$ & $0.293 \pm 0.192$ & $0.718\pm 0.409$ \\
    \citet{ONeill2021} (Raw) & 222 & 277 & 11.8 & 4.56 & $0.419 \pm 0.067$ & $0.021 \pm 0.119$ & $0.907 \pm 0.176$ \\
    **\citet{Cheng2018} (True) & 158 & 100 & 2.00 & 0.92 & $1.239 \pm 0.172$ & $1.161 \pm 0.247$ & $1.099\pm 0.834$ \\
    **\citet{Liu2018} (True) & 275 & 159 & 3.30 & 1.28 & $0.860 \pm 0.106$ & $0.979 \pm 0.167$ & $0.837\pm 0.397$ \\
    **\citet{ONeill2021} (True) & 614 & 251 & 4.72 & 1.13 & $0.537 \pm 0.062$ & $0.392 \pm 0.120$ & $0.919 \pm 0.151$ \\
    \hline
    \end{tabular}
    \bigskip
    \\
    \raggedright
    Notes. *: For the case with density threshold applied, the cores are still selected from the mass surface density map without density threshold, but then cores with zero mass are removed. **:For the ``True'' CMFs, core numbers and statistical properties (maximum, arithmetic mean and geometric mean) are derived from the core mass functions by assuming that all cores in a bin have the same mass as the center of the bin \citep[see][]{Cheng2018, Liu2018, ONeill2021}.
\end{table*}

First considering the high-mass end of the CMF, i.e., $M\geq 10\:M_\odot$, in the colliding case at 2, 3 and 4~Myr, we find $\alpha_{10} = 0.779\pm0.289$, $0.585 \pm 0.091$ and $0.693 \pm 0.051$. If ALMA filtering is applied, these numbers change to $\alpha_{10} = 0.667 \pm 0.247$ and $0.678 \pm 0.073$ for the cases of 3 and 4~Myr that have sufficient numbers of cores for this analysis. If a density threshold of $n_{\rm H} = 10^4$ or $10^5\:{\rm cm}^{-3}$ is applied when assessing core mass in the non-ALMA-filtered images at 4~Myr, then we find $\alpha_{10}=0.791\pm0.055$ and $0.763\pm0.061$, respectively. Thus we find that these results for the high-end CMF index are fairly insensitive to these various methods and the derived high-end power law index is shallower (i.e., more top-heavy) than the Salpeter index.

For the same high-end mass range in the non-colliding case at 4~Myr we find $\alpha_{10} = 0.784\pm0.144$ in the original simulation data, $1.111\pm0.352$ after ALMA-filtering, and $1.068\pm0.179$ and $1.180\pm0.265$ for the density thresholds of $n_{\rm H} = 10^4$ or $10^5\:{\rm cm}^{-3}$. Especially after ALMA-filtering or applying a density threshold, we find the high-end CMF in the non-colliding case has a steeper index (i.e., fewer massive cores) than the colliding case and is closer to the Salpeter index.

The above simulation results can be compared to observed CMFs: e.g., in IRDC clumps by \citet{Liu2018}, whose data imply a ``raw'' CMF index $\alpha_{10}=0.718\pm0.409$ and a ``true'' CMF index $\alpha_{10}=0.837\pm0.397$ (i.e., after flux and number completeness corrections); in massive clumps by \citet{ONeill2021}, whose data imply $\alpha_{10}=0.907\pm0.176$ (raw) and $\alpha_{10}=0.919\pm0.151$ (true). Direct comparison with individual regions, e.g., the study of G286 by \citet{Cheng2018}, in this mass regime is typically hampered by the relatively small numbers of cores leading to large uncertainties in the derived power law index. However, overall for the high-mass end of the CMF we find consistency in our simulations results with the observational results from the multi-region samples of \citet{Liu2018} and \citet{ONeill2021}. However, given the state of the observational uncertainties and the relatively limited number of cores in this mass range in the simulated clouds (especially the non-colliding case), we are not able to use the results to conclude if the colliding or non-colliding results are a better match to the observed systems.

We next consider fits to the mass range $M\geq 1\:M_\odot$. Inspecting these fits that are shown in Figure~\ref{fig:cmf-sim-alma-thres} we see that the CMF distributions are often not particularly well described by a single power law. It is the ALMA-filtered CMFs that appear to be best described by a single power law over this full mass range. Thus, the original simulation results without ALMA filtering yield very shallow values of $\alpha_1 \sim 0.2$ to 0.4. Application of a density threshold causes a slight steepening of this index. The ALMA-filtered CMF index has values of $\alpha_1=0.546\pm0.066$ and $0.440\pm0.030$ in the colliding case at 3 and 4~Myr and $0.492\pm0.074$ in the non-colliding case at 4~Myr. The observational results over this mass range are much steeper in the case of G286 \citep{Cheng2018}. For IRDC clumps the initial ``raw'' estimate before flux and completeness corrections has a value of $\alpha_1 = 0.495 \pm0.100$ \citep{Liu2018}, while for massive clumps it is $0.419\pm0.067$ \citep{ONeill2021}. We thus see that, similar to the case for $M\geq10\:M_\odot$, our simulation results are consistent with observational measures of the CMF for $M\geq1\:M_\odot$. However, again, it does not appear possible to distinguish between the colliding and non-colliding case via this metric.

\begin{figure}
    \centering
    \includegraphics[width=\linewidth]{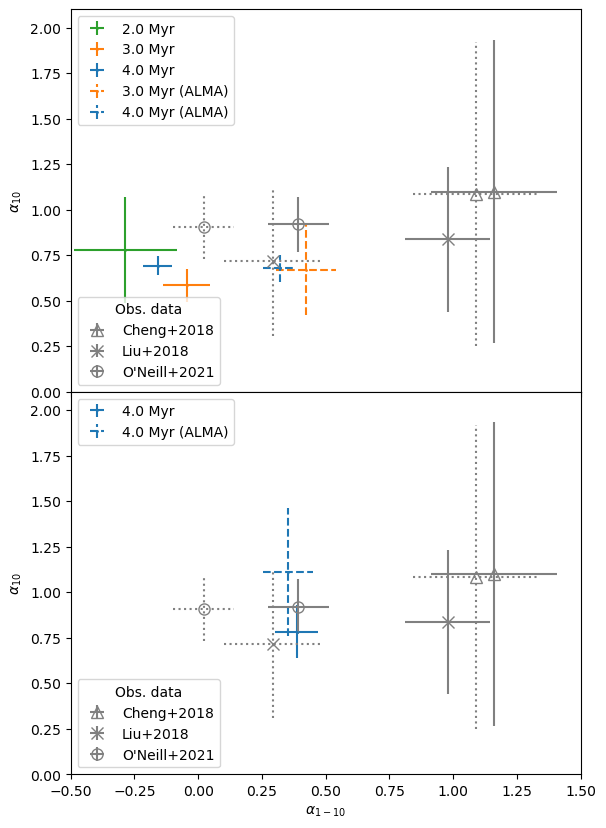}
    \caption{Scatter plots of $\alpha_{10}$ versus $\alpha_{1-10}$. {\it (a) Top panel:} Original and ALMA-filtered GMC collision simulation data are shown with solid and dashed error bars, respectively, with colors distinguishing different times. The power law indices derived from ``raw'' and ``true'' CMFs from observations are plotted with grey dotted and solid error bars, respectively. 
    {\it (b) Bottom panel:} As (a), but for the non-colliding simulation.
    }
    \label{fig:alpha-corel}
\end{figure}

Finally, we consider the CMF power law index when fit only to the range $1\leq M/M_\odot \leq 10$. A comparison in this limited mass range may be important as \citet{ONeill2021} found evidence for a break in power law behaviour of the CMF at $\sim 10\:M_\odot$. The colliding case before ALMA filtering yields values of $\alpha_{1-10} = -0.044\pm0.092$ and $-0.160\pm0.055$ at 3 and 4~Myr, i.e., a rising function with mass. After ALMA filtering these values become $\alpha_{1-10}=0.423\pm0.117$ and $0.320\pm0.065$. The non-colliding case before ALMA filtering is significantly steeper than the corresponding colliding case, i.e., with $\alpha_{1-10}=0.386\pm0.084$. ALMA filtering hardly changes this value, i.e., it becomes $\alpha_{1-10}=0.353\pm0.097$. The observational results (from raw CMFs) in this mass range are $\alpha_{1-10}=1.09\pm0.25$ in G286 \citep{Cheng2018}, $0.293\pm0.192$ in IRDC clumps \citep{Liu2018} and $0.021\pm0.119$ in massive clumps \citep{ONeill2021}. We see that our simulation results can match CMF properties in IRDCs, but not in G286 and massive clumps. To summarize the above results, in Figure~\ref{fig:alpha-corel} we show a diagram of $\alpha_{1-10}$ versus $\alpha_{10}$.

\begin{figure*}
    \centering
    \includegraphics[width=\linewidth]{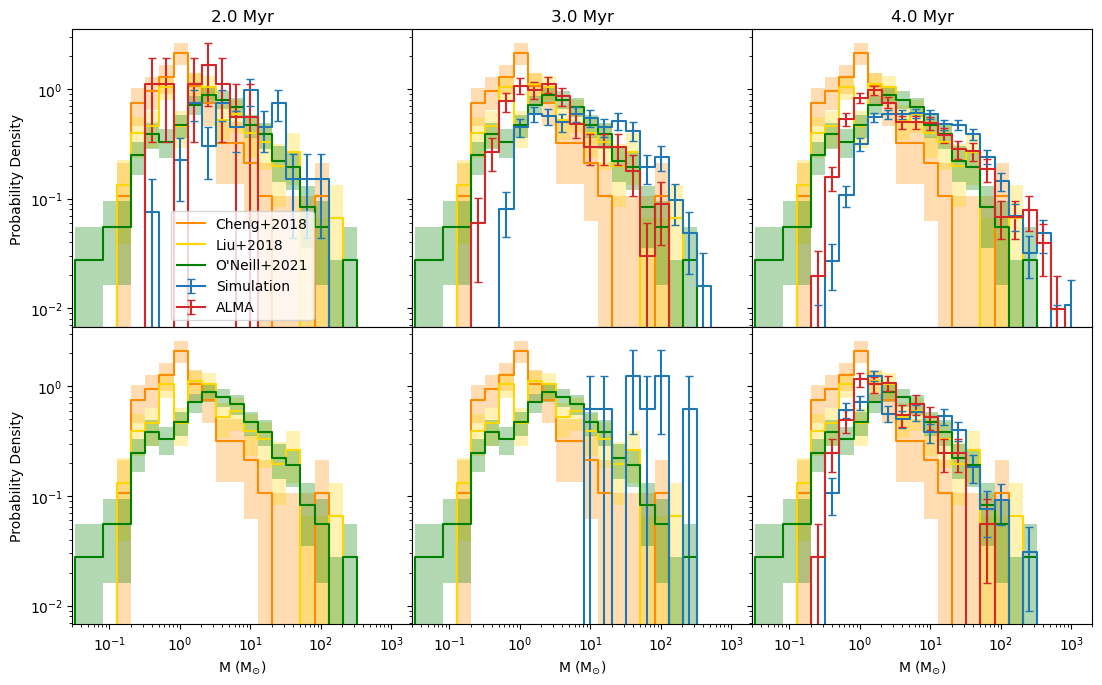}
    \includegraphics[width=\linewidth]{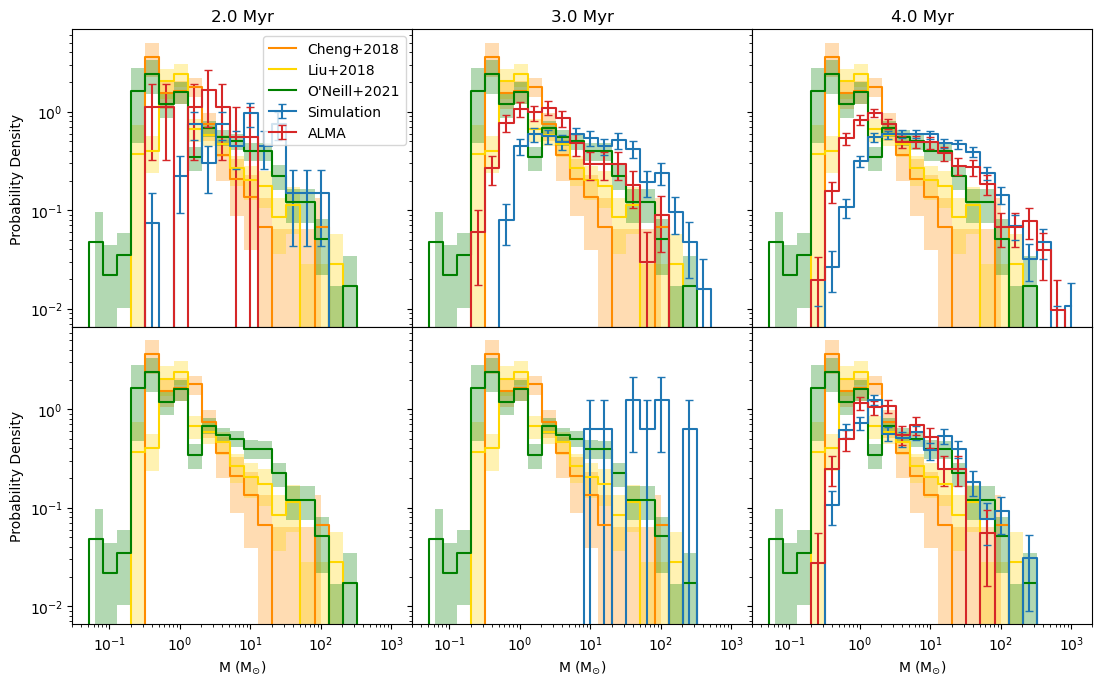}
    \caption{
    The probability distribution function (PDF) of core mass functions (CMFs) at 2~Myr (left column), 3~Myr (middle column) and 4~Myr (right column) for the colliding case (each first row) and the non-colliding case (each second row). The blue lines show original simulation results and the red lines show those after ALMA filtering. The error bars indicate the Poisson counting errors. The top set of six panels show comparison to observational ``raw'' CMFs from \citet{Cheng2018, Liu2018} and \citet{ONeill2021}. The bottom set of six panels show the observational ``true'' CMFs, i.e., after flux and number correction, from these studies.
    }
    \label{fig:cmfpdf}
\end{figure*}

For a more complete comparison with the observational CMFs
we plot the probability density of the CMFs in Figure~\ref{fig:cmfpdf}. The CMFs are normalized by the number of cores whose masses are $\geq 1 M_\odot$, i.e., to avoid the uncertainties from the lowest-mass cores. The top set of panels shows ``raw'' CMFs, while the bottom set show ``true'' CMFs (i.e., after flux and number completeness corrections have been applied).

We next compare the simulated and observed CMF PDFs via the Kolmogorov–Smirnov (KS) test. We set a lower bound of the CMFs to reduce the influence of low-mass cores. According to the clipped CMFs, we generate random samples in each bin to obtain $p$-values by {\em ks\_2samp} in the {\em scipy} package. The final $p$-value of each comparison is then calculated by the mean value of 3,000 bootstrap resamplings. As the small cores have higher uncertainties in the observed samples, we set the lower limit of the range as being the mass bin centered at $2.51\:M_\odot$. 

\begin{figure*}
    \centering
    \includegraphics[width=\linewidth]{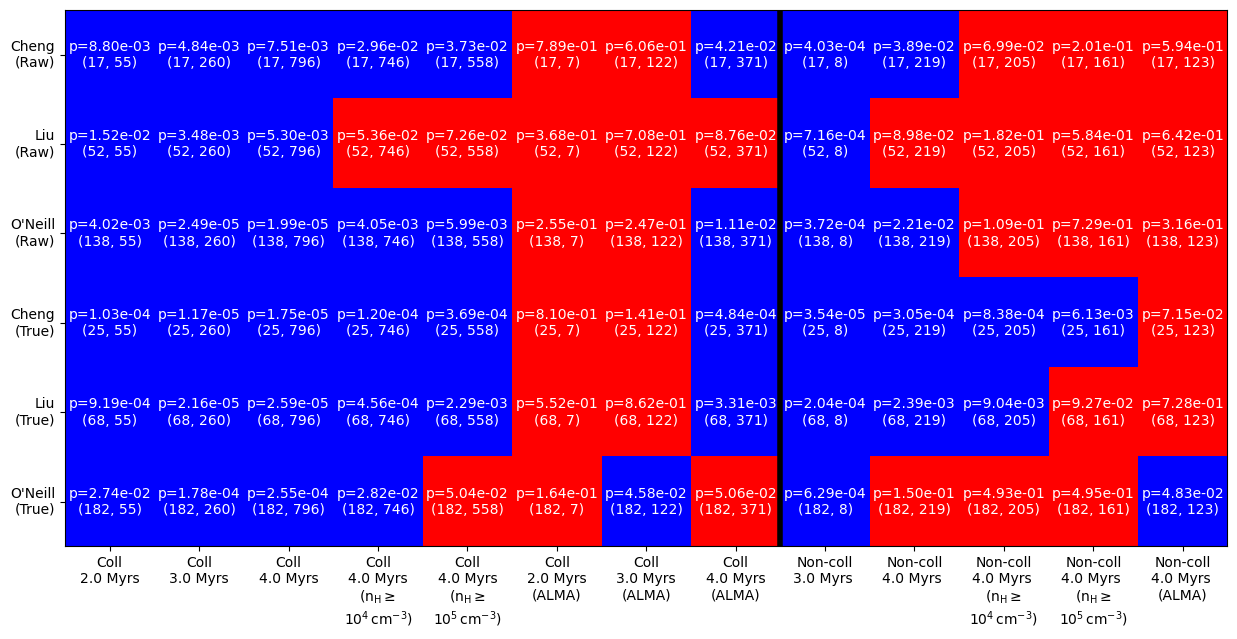}
    \caption{KS test $p$-values by comparing simulation results with the ``raw'' and ``true'' CMFs derived from observational studies. Blue or red background colors indicate whether the null hypothesis is rejected or not, respectively, i.e., a blue color indicates a significant difference between the simulation and the observation. The numbers in the parentheses indicate the effective core numbers in the observations and simulations.}
    \label{fig:kstest_all}
\end{figure*}

In Figure~\ref{fig:kstest_all}, we display the $p$-values of the KS tests by comparing the simulation results against the observed CMFs. A panel is colored red if the null hypothesis is not rejected ($p > 0.05$), i.e., the two distributions may come from the same population. Otherwise, we color panels in blue. These results show that there is consistency in the distributions especially when our ALMA filtered results are compared to the observed ``raw'' CMFs. In the colliding case, there is a  modest preference to favor the results from intermediate times, i.e., $t=3\:$Myr, over those from the final time at 4~Myr. Note, one must be aware of the effects of small numbers of cores, which makes it easier to achieve consistency: this is especially the case for the ALMA-filtered colliding case at 2~Myr. In general, similar to what we found with the comparison of power law indices, examples of both colliding and non-colliding cases of the ALMA-filtered simulations are consistent with the observed CMFs.

\subsection{Core Sizes and Densities}

Moving beyond the mass function, we next examine the intrinsic physical properties of the identified cores. In the previous sections, we examined cores and their masses defined in multiple ways. Here, we fix dendrogram parameters to our fiducial case, and examine the properties of the cores identified in the original simulated $\Sigma$ map, as well as in the ALMA synthetic observation. Then we compare the results with observed core properties.

\begin{figure*}
    \centering
    \includegraphics[width=\linewidth]{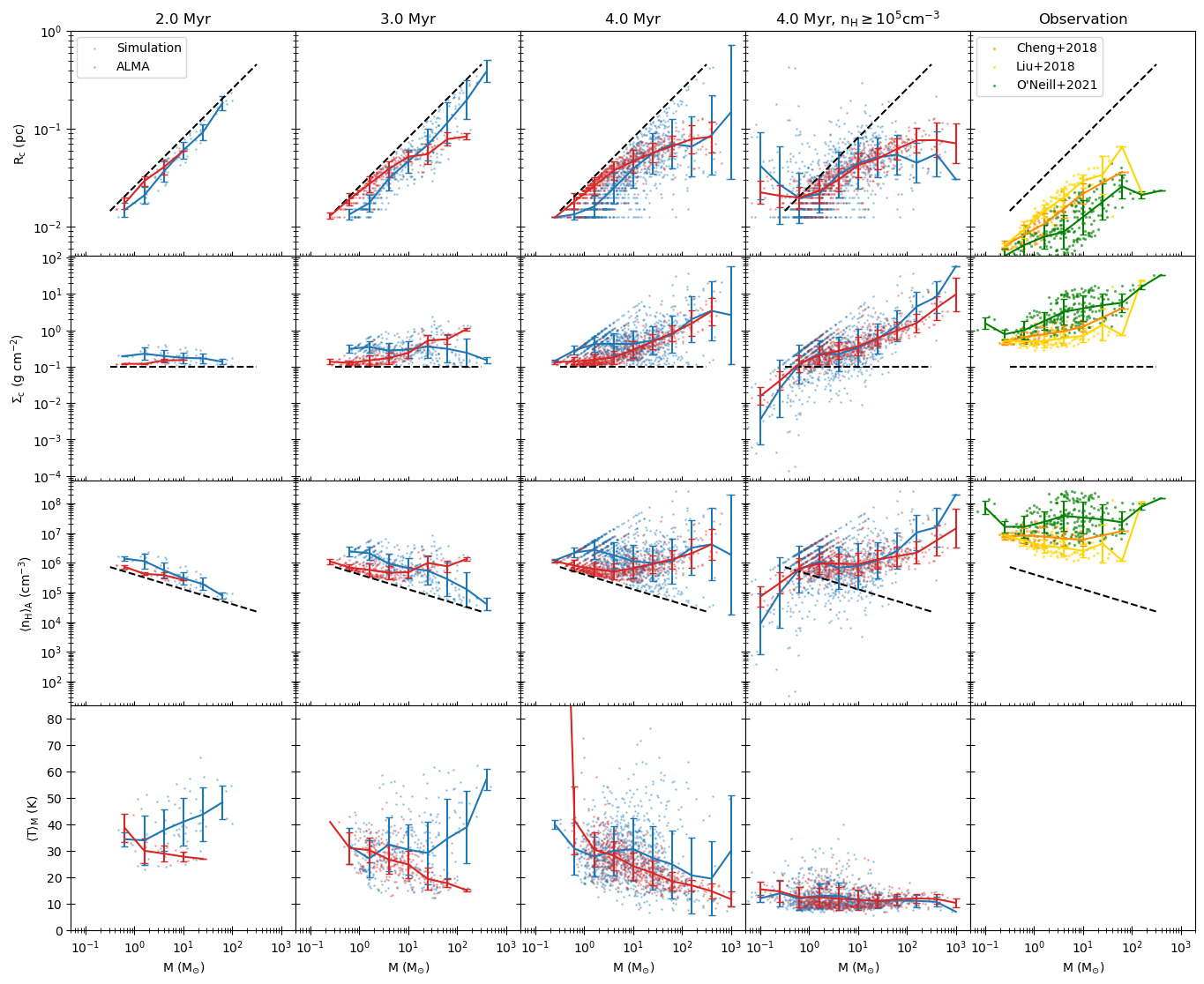}
    \caption{Core properties determined in projection for the colliding case. The columns show results at 2~Myr (first column), 3~Myr (second column), 4~Myr (third column), after applying density threshold of $n_{\rm H} \geq 10^{5}\:{\rm cm}^{-3}$ at 4~Myr (fourth column) and the observational results (fifth column). {\em Top Row:} Effective radius $R_{c} = \sqrt{A_{c}/\pi}$. {\em Second Row:} Mean mass surface density $\Sigma_c = M_c / A_c$. {\em Third Row:} Mean number density (see text). {\em Bottom Row:} Mass-weighted temperature. Dashed lines show limits based on the requirement that $\Sigma \geq 0.1 {\rm g cm}^{-3}$. The blue/red lines and their error bars indicate the mean values and the standard deviations in each bin, whose width is set to be 0.4 dex. The observational data from \citet{Cheng2018, Liu2018} and \citet{ONeill2021} are plotted in orange, gold and green, in the last column (note there are no observational constraints on temperature). For the first three rows, the means and standard deviations are shown for the logarithmic quantities, while in the fourth row these are done in linear space.
    }
    \label{fig:coreprops_col}
\end{figure*}

In the first row of Figure~\ref{fig:coreprops_col}, for the colliding case we plot the effective radii of the cores $R_{c} \equiv \sqrt{A_{c}/\pi}$, where $A_{c}$ is the projected area of cores. The results of the original simulation and those based on synthetic ALMA observation are plotted in blue and red, respectively. For a core with certain mass, the black dashed line indicates its maximum radius with the assumption of $\Sigma_c = 0.1\:{\rm g \:cm}^{-2}$. We plot the average mass surface densities of the cores in the second row. Initially, for both original and ALMA-filtered cases, the cores demonstrate limited variation in mass surface density, with average values only slightly above the threshold. Therefore, the core radii are also close to their maximum values. As time evolves, the average mass surface densities gradually show a positive correlation with mass. 
If we examine only contributions from high-density gas (with a density threshold $n_{\rm H} \geq 10^{5}\: {\rm cm}^{-3}$) in the final states (the fourth column), this correlation appears stronger.

\begin{figure*}
    \centering
    \includegraphics[width=\linewidth]{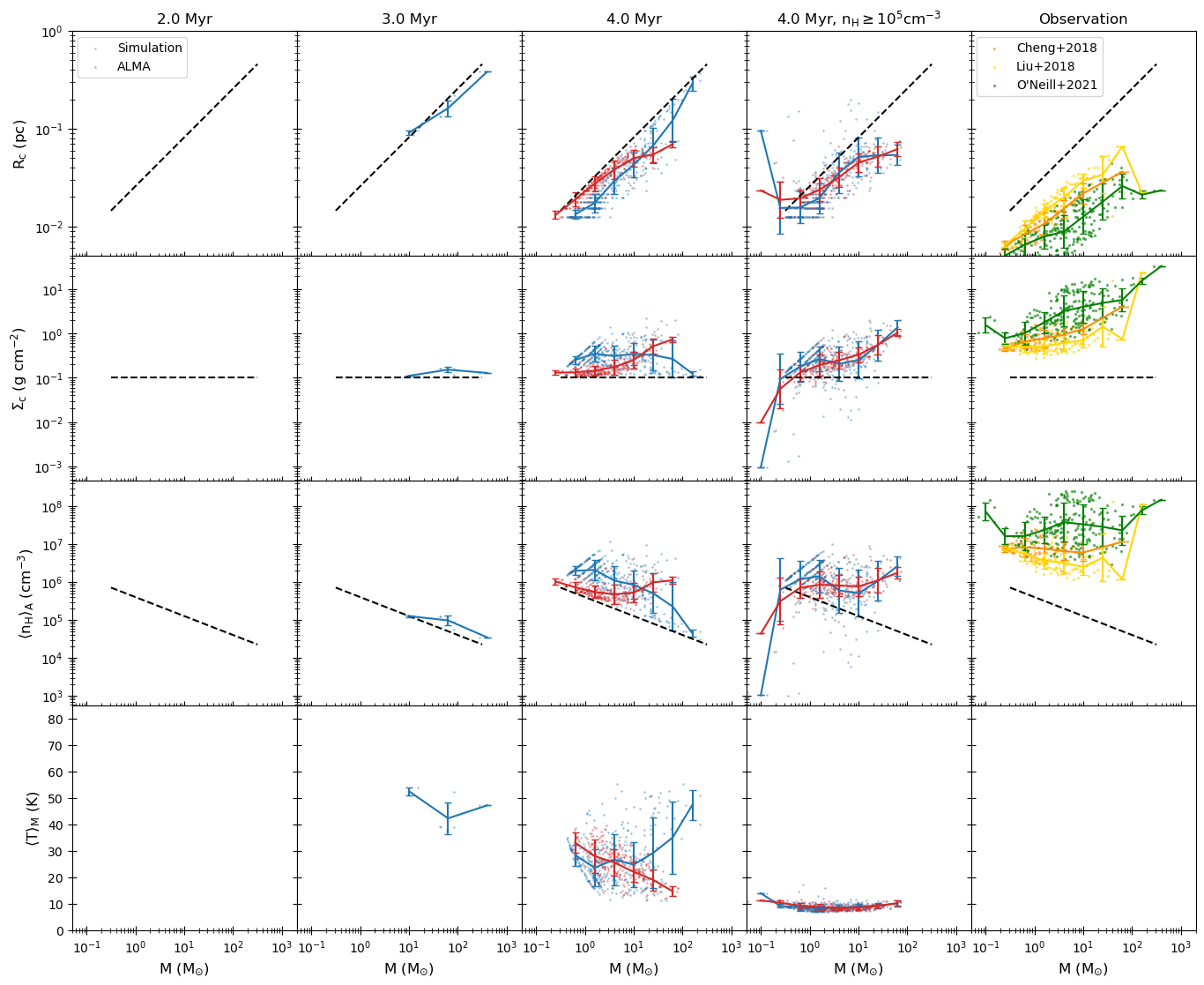}
    \caption{As Fig.~\ref{fig:coreprops_col}, but now for the non-colliding case. 
    }
    \label{fig:coreprops_nocol}
\end{figure*}

Since the cores are defined via projection, the lack of information about the third dimension causes difficulties in estimating the density. Therefore, we estimate the number densities of the cores given an assumed spherical geometry, i.e., via:
\begin{equation}
\langle n_{\rm H} \rangle_A = \frac{3 M_c}{\mu_{\rm H} 4\pi R_c^3} = \frac{3\pi^{1/2} M_c}{4 \mu_{\rm H} A_c^{3/2}},
\label{eqn:densarea}
\end{equation}
where we adopt a mass per H of $2.34\times 10^{-24}\:$g (assuming $n_{\rm He}=0.1 n_{\rm H}$ and ignoring other species). The derived volume densities $\langle n_{\rm H} \rangle_A$ of the colliding case are shown in the third row of Figure~\ref{fig:coreprops_col}. Given that cores are defined by a mass surface density threshold and the assumption that the volume is $V = 4\pi {R_c}^3/3$, there is a minimum volume density that varies inversely with core mass ($n_{\rm H,min} \propto M_c^{-1/2}$). We see that at early times, the derived volume densities are close to this minimum. However, we find that by 4~Myr, $\langle n_{\rm H} \rangle_A$ no longer closely follows the minimum, but tends to increase for more massive cores.
Furthermore, if we apply a density threshold ($n_{\rm H} \geq 10^5\:{\rm cm}^{-3}$) to define the core material, then this trend is enhanced.



In Figure~\ref{fig:coreprops_nocol} we plot the same above properties of the cores, but now for the non-colliding case.
Since the cores develop more slowly than the colliding case, the core radii and mean mass surface densities still closely follow the maximum radius and minimum mass surface density at 4~Myr. The volume density then shows the corresponding behavior implied by this limit. Although the synthetic ALMA observation reduces the radii of the most massive cores and increases the densities, the correlation between density and mass remains quite weak. Similarly, applying a density threshold ($n_{\rm H} \geq 10^5\:{\rm cm}^{-3}$) increases the densities of the cores, but the density versus mass relation remains quite flat.

Comparing with the observational data for these quantities, we see that our simulated cores tend to have larger radii and thus lower densities. One potential cause of this is the observed regions are typically closer than our adopted fiducial distance of 5~kpc and, as mentioned above, the ALMA observations are thus typically able to resolve smaller scales that we probe in the simulations. Simulations with higher spatial resolution are needed to assess this aspect. However, another potential issue is that the observed cores are already protostellar sources, i.e., with a significant protostellar mass and associated heating that has an effect of concentrating mm continuum flux that is used in the observational definition of the sizes. To address this aspect, one could either focus on a sample of pre-stellar cores that are selected from mass surface density maps, or one could implement sub-grid models of protostellar cores in the simulations that induce local heating and associated enhanced mm flux emission. We defer such steps for future work, but discuss these aspects further in \S\ref{sec:discussion}.



\begin{figure}
    \centering
    \includegraphics[width=\columnwidth]{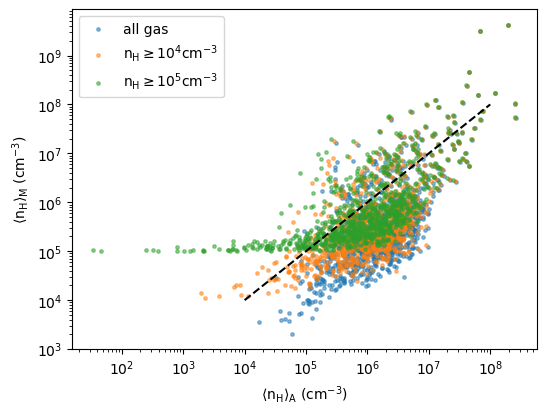}
    \caption{Comparison of different methods for estimating the core number densities. Along the $x$-axis is the number density as estimated from the projected area, as in Equation~\ref{eqn:densarea}. The $y$-axis shows the number density estimated from the mass-weighted density, as in Equation~\ref{Eqn:MassWeight}. The dashed line shows the one to one relation.}
    \label{fig:denscomp}
\end{figure}

We also compute $\langle n_{\rm H} \rangle_M$ for each core in the colliding case. This provides a different estimate of the density without introducing assumptions about the core geometry along the line of sight. To reduce the contribution of low-density gas along the line of sight, we also consider cases with density thresholds $n_{\rm H} \geq 10^4\:{\rm cm}^{-3}$ and $n_{\rm H} \geq 10^5\:{\rm cm}^{-3}$. Figure~\ref{fig:denscomp} compares $\langle n_{\rm H} \rangle_A$ and these mass-weighted densities. In the case without any density threshold, the densities derived from the projected area tend to one order of magnitude higher than those derived from the full integration along the line of sight. As a density threshold is applied, the mass-weighted densities become closer to $\langle n_{\rm H} \rangle_A$, except for the cores whose density is smaller than $10^4\:{\rm cm}^{-3}$. Overall, applying a density threshold of $10^5\:{\rm cm}^{-3}$ yields a better agreement rather than that from $10^4\:{\rm cm}^{-3}$.

\subsection{Core Temperatures}\label{sec:temp}

We calculate the mass-weighted temperatures $\langle T\rangle_M$ of the cores and show the results in the bottom rows of Figures~\ref{fig:coreprops_col} and \ref{fig:coreprops_nocol} for the colliding and non-colliding cases, respectively. For cores selected from the original simulation data, i.e., no ALMA filtering or density threshold applied, in the colliding case, the core temperatures initially have a weak dependency on core mass. The temperatures range from 20~K to 60~K at 2~Myr, with more massive cores tending to have higher temperatures. As time evolves to 3~Myr, more and more cores become cooler, except for some rare examples of cores hotter than $50\:$K.
However, most cores are cooler than 40~K by 4~Myr. The overall mean temperatures are 44.8~K, 36.4~K and 24.7~K at 2, 3, and 4~Myr. In the non-colliding case, the cores show similar behavior, i.e., becoming cooler as time evolves, but the massive cores still have temperatures $>40$~K at 4~Myr. 

\begin{figure}
    \centering
    \includegraphics[width=\columnwidth]{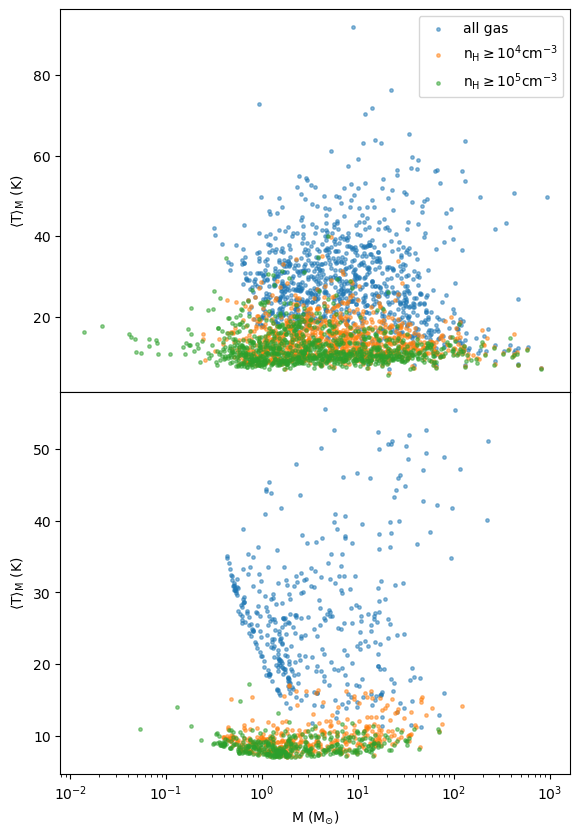}
    \caption{Core temperatures $\langle T\rangle_{M}$ for the colliding case (top panel) and the non-colliding case (bottom panel) at $t=4$~Myr. Cores are identified from all gas in the projected image. Blue circles indicate the average core temperature including all gas. Orange circles indicate average core temperature contributed by cells where $n_{\rm H} \geq 10^4\,{\rm cm^{-3}}$. Green circles indicate average core temperature contributed by cells where $n_{\rm H} \geq 10^5\,{\rm cm^{-3}}$.}
    \label{fig:tempcomp}
\end{figure}

However, cores identified in synthetic ALMA observation, i.e., from the ALMA-filtered images, show a different behavior. Now the most massive cores are cooler. This reflects the dramatically effects of ALMA filtering on defining cores.
Application of density thresholds also has a large impact on derived temperatures. 
In Figure~\ref{fig:tempcomp}, we see that the core temperatures drop to around 20~K if a density threshold $n_{\rm H} \geq 10^4\:{\rm cm}^{-3}$ is applied. If applying a higher density threshold of $n_{\rm H} \geq 10^5\:{\rm cm}^{-3}$, then the cores have even cooler temperatures, $\sim 10$~K. 

As a comparison, we list here the mean temperatures of cores in the three cases: all gas; $n_{\rm H} \geq 10^4\:{\rm cm}^{-3}$; and $n_{\rm H} \geq 10^5\:{\rm cm}^{-3}$ at 4~Myr. For these the mean ($\pm$ dispersion) core temperatures are $24.7 (\pm 14.3)$~K, $12.6 (\pm 3.6)$~K and $11.1 (\pm 2.8)$~K, respectively.
For the non-colliding case, we find $32.6 (\pm 13.6)$~K, $11.0 (\pm 2.3)$~K and $9.1 (\pm 1.2)$~K for these cases, respectively.
We note the possibility of increased rates of adiabatic heating for cores that are forming more rapidly, which is likely to the be the case in the cloud collision simulation. New observations, e.g., via high resolution $\rm NH_3$ observations, are needed to derive temperatures in the observed regions of \citet{Cheng2018,Liu2018} and \citet{ONeill2021} in order to make direct comparisons with the core populations that we have considered for the CMFs.



\subsection{Core Virial Parameters}

\begin{figure}
    \centering
    \includegraphics[width=\columnwidth]{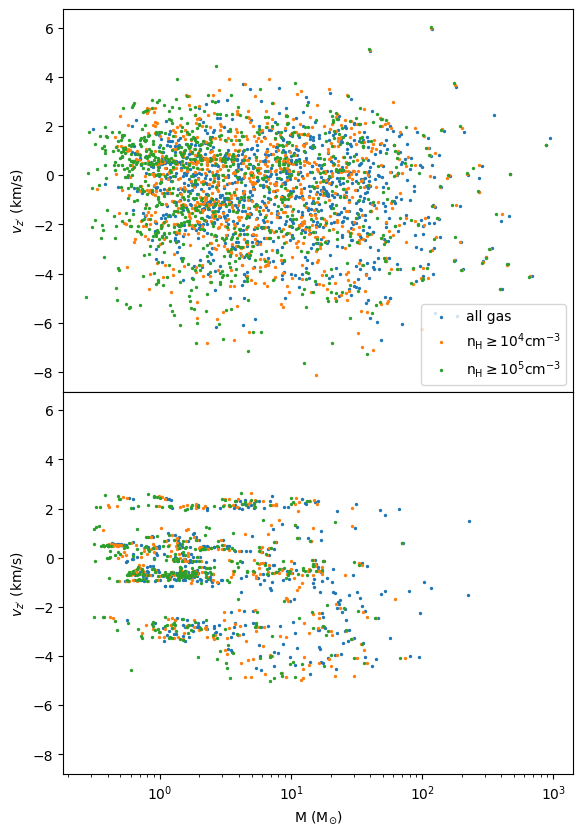}
    \caption{Scatter plots of core radial velocity and mass at 4~Myr for the colliding (top panel) and non-colliding (bottom panel) cases. Blue dots show the velocity from considering all gas along the line of sight. Orange and green dots only calculate the velocity from the material with $n_{\rm H} \geq 10^4\:{\rm cm}^{-3}$ and $n_{\rm H} \geq 10^5\:{\rm cm}^{-3}$, respectively.}
    \label{fig:vrad}
\end{figure}

\begin{figure}
    \centering
    \includegraphics[width=\columnwidth]{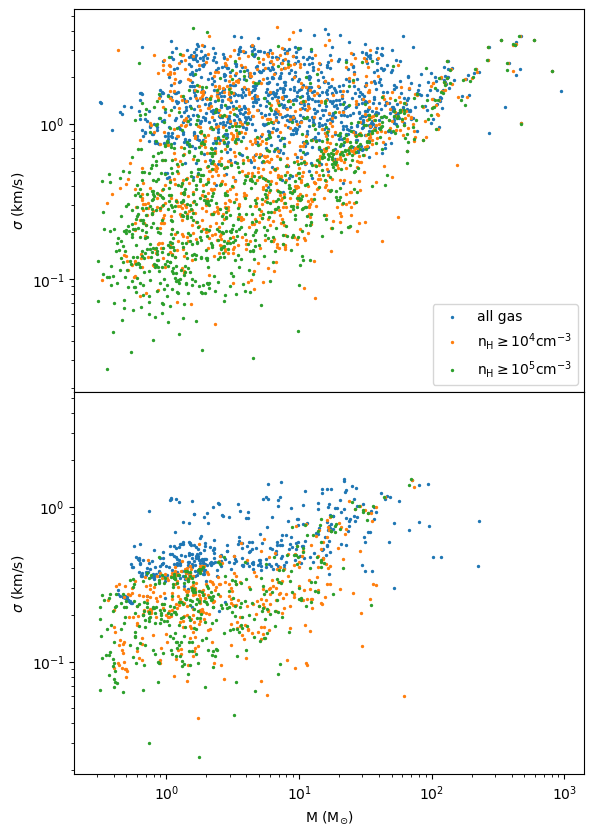}
    \caption{Similar to Figure~\ref{fig:vrad}, but now showing core velocity dispersion versus mass.}
    \label{fig:vdisp}
\end{figure}

To estimate the gravitational boundedness of the cores, we calculate the virial parameter \citep{1992ApJ...395..140B}
\begin{equation}
\alpha_{\rm vir} = 5 \sigma^2 R_c / (G M),
\label{eqn:virial}
\end{equation}
where $\sigma$ is the one-dimensional velocity dispersion. The velocity dispersion is estimated by the standard deviation of the mass-weighted line-of-sight velocity in each core. Figure~\ref{fig:vrad} shows the radial velocity of each core and Figure~\ref{fig:vdisp} shows the velocity dispersion and the mass of the cores. We show a scatter plot for three cases: all gas is included; only gas with $n_{\rm H} \geq 10^4\:{\rm cm}^{-3}$ is included; and only gas with $n_{\rm H} \geq 10^5\:{\rm cm}^{-3}$ is included. 

In Figure~\ref{fig:vdisp}, we see the velocity dispersions generally have higher values in the colliding case. As the density threshold is applied, the values shift to lower levels. The velocity dispersion can be about 10 times smaller when a threshold of $n_{\rm H} \geq 10^5\:{\rm cm}^{-3}$ is applied, especially for low-mass cores.

\begin{figure}
    \centering
    \includegraphics[width=\columnwidth]{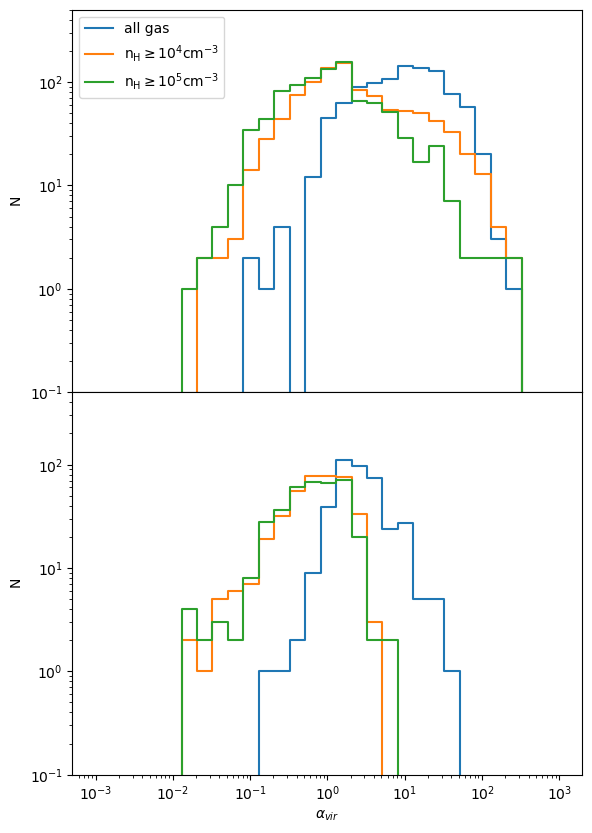}
    \caption{Distribution of virial parameters $\alpha_{\rm vir}$ for the colliding case (top) and the non-colliding case (bottom).}
    \label{fig:virial}
\end{figure}



Figure~\ref{fig:virial} shows the distribution of the virial parameters for the cores, based on the velocity dispersion measurements shown in Figure~\ref{fig:vdisp}. Since the velocity dispersions in the colliding simulation are higher, we see larger virial parameters in this case. Applying a density threshold tends to reduce virial parameter values, with the peak being closer to unity. Nevertheless, we see the cores have a broad range of virial parameters.
In the colliding case, the mean ($\pm$ dispersion) values of $\log{\alpha_{\rm vir}}$ are $0.96 (\pm 0.54)$, $0.34 (\pm 0.70)$ and $0.034 (\pm 0.74)$ for the cases of all gas, $n_{\rm H} \geq 10^4\:{\rm cm}^{-3}$, and $n_{\rm H} \geq 10^5\:{\rm cm}^{-3}$, respectively. In the non-colliding case, the corresponding values are $\log{\alpha_{\rm vir}}=0.41 (\pm 0.33)$, $-0.17 (\pm 0.41)$ and $-0.22 (\pm 0.42)$. For reference, each of the initial clouds has a velocity dispersion of $\sigma = 5.2\,{\rm km\, s^{-1}}$ and a virial parameter $\alpha_{\rm vir} = 6.8$.

In Figure~\ref{fig:virial}, we see that most cores are supervirial if no density threshold is applied. In the colliding case, around half of the cores are still supervirial after application of a density threshold. In contrast, more cores are subvirial after a density threshold is applied in the non-colliding case. To be more specific, for the colliding case at 4 Myr there are initially only 38 subvirial cores out of total 984 cores, i.e., 3.9\%. In the case of density thresholds of $n_{\rm H} \geq 10^4\:{\rm cm}^{-3}$ and $n_{\rm H} \geq 10^5\:{\rm cm}^{-3}$, the fractions of subvirial cores increase to 330/983 (33.6\%) and 454/933 (48.7\%), respectively. If we further check the fraction of gravitationally bounded ($\alpha_{\rm vir} < 2$) cores, these are 126/984 (12.8\%), 558/983 (56.8\%) and 669/933 (71.7\%) in the three cases. For the non-colliding simulation, the fractions of subvirial cores are 23/395 (5.8\%), 250/395 (63.3\%) and 251/372 (67.5\%) and of bound cores are 164/395 (41.5\%), 359/395 (90.9\%) and 348/372 (93.5\%) for these three density threshold cases. The fraction of unbound ($\alpha_{\rm vir}>2$) cores selected with the density threshold of $n_{\rm H} \geq 10^5\:{\rm cm}^{-3}$ is most sensitive to whether (28\%) or not (6.5\%) the cores formed from a GMC-GMC collision. Thus we see that a survey of the dynamical state of cores has the potential to distinguish between colliding and non-colliding formation scenarios.



\begin{figure*}
    \centering
    \includegraphics[width=0.95\linewidth]{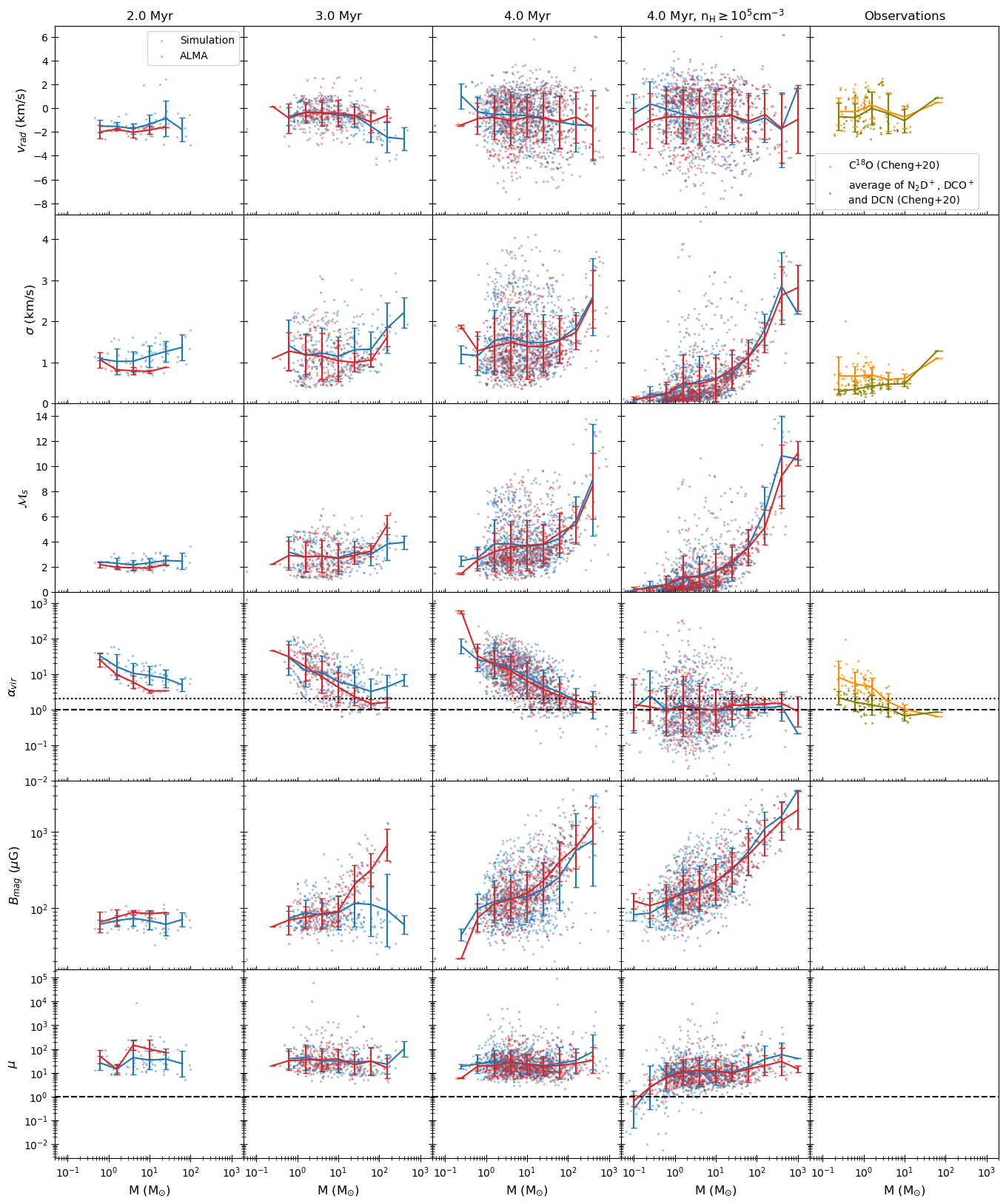}
    \caption{The kinematic, dynamic and magnetic properties of cores in the colliding case and compared to observed cores. From left to right: 2~Myr, 3~Myr, 4~Myr, 4~Myr data with density threshold $n_{\rm H} \geq 10^5\:{\rm cm}^{-3}$, and observational data. {\it Top Row}: Radial velocity. {\it Second Row}: 1-D velocity dispersion. {\it Third Row}: Mach number based on the sound speed derived from mean core temperature. {\it Fourth Row}: Virial parameter. Dashed line and dotted line indicate $\alpha_{\rm vir} = 1$ (virial equilibrium) and $\alpha_{\rm vir}=2$ (gravitationally bound). {\it Fifth Row}: Magnetic field strength weighted by density. {\it Sixth Row}: Mass-to-flux ratio. In each panel of the first four columns, blue dots show cores found in the original simulation data and red dots show cores identified in the ALMA filtered data. The solid line shows the mean value and the error bar shows the standard deviation. The bins are defined the same as in Figure~\ref{fig:coreprops_col}. The mean and standard deviation of radial velocity and velocity dispersion are weighted by core mass. Virial parameter, magnetic field strength, and mass-to-flux ratio show the mean and standard deviation of their logarithmic values.}
    \label{fig:corekin_col}
\end{figure*}

In Figure~\ref{fig:corekin_col}, we further analyze the kinematics of the cores in the colliding case. We follow the same columns and colors for original and ALMA-filtered cases as in Figures~\ref{fig:coreprops_col} and \ref{fig:coreprops_nocol}. In the first row, we plot the radial velocities of the cores. The second row shows the one-dimensional velocity dispersion. We find that the velocity dispersion globally becomes larger due to the influence of ambient gas. If we remove the contribution of ambient gas, most small cores have small velocity dispersions, $< 1\:$km/s, but massive cores still retain high values. With the velocity dispersion and the temperature in Figure~\ref{fig:coreprops_col}, the Mach number $\mathcal{M}_s \equiv \sigma/c_s$, where $c_s = \sqrt{\gamma kT/\mu}$ is the sound speed at that temperature, is plotted in the third row. Mach numbers show similar behavior as the velocity dispersion and more massive cores have higher values, especially when the density threshold is applied. 
The fourth row shows the virial parameter based on the 1-D velocity dispersion. For this colliding case, most cores are supervirial and only a few are subvirial. The result does not change if cores are defined after ALMA filtering. However, if the density threshold $n_{\rm H} \geq 10^5\: {\rm cm}^{-3}$ is applied, about half of cores become subvirial. 

\begin{figure*}
    \centering
    \includegraphics[width=0.95\linewidth]{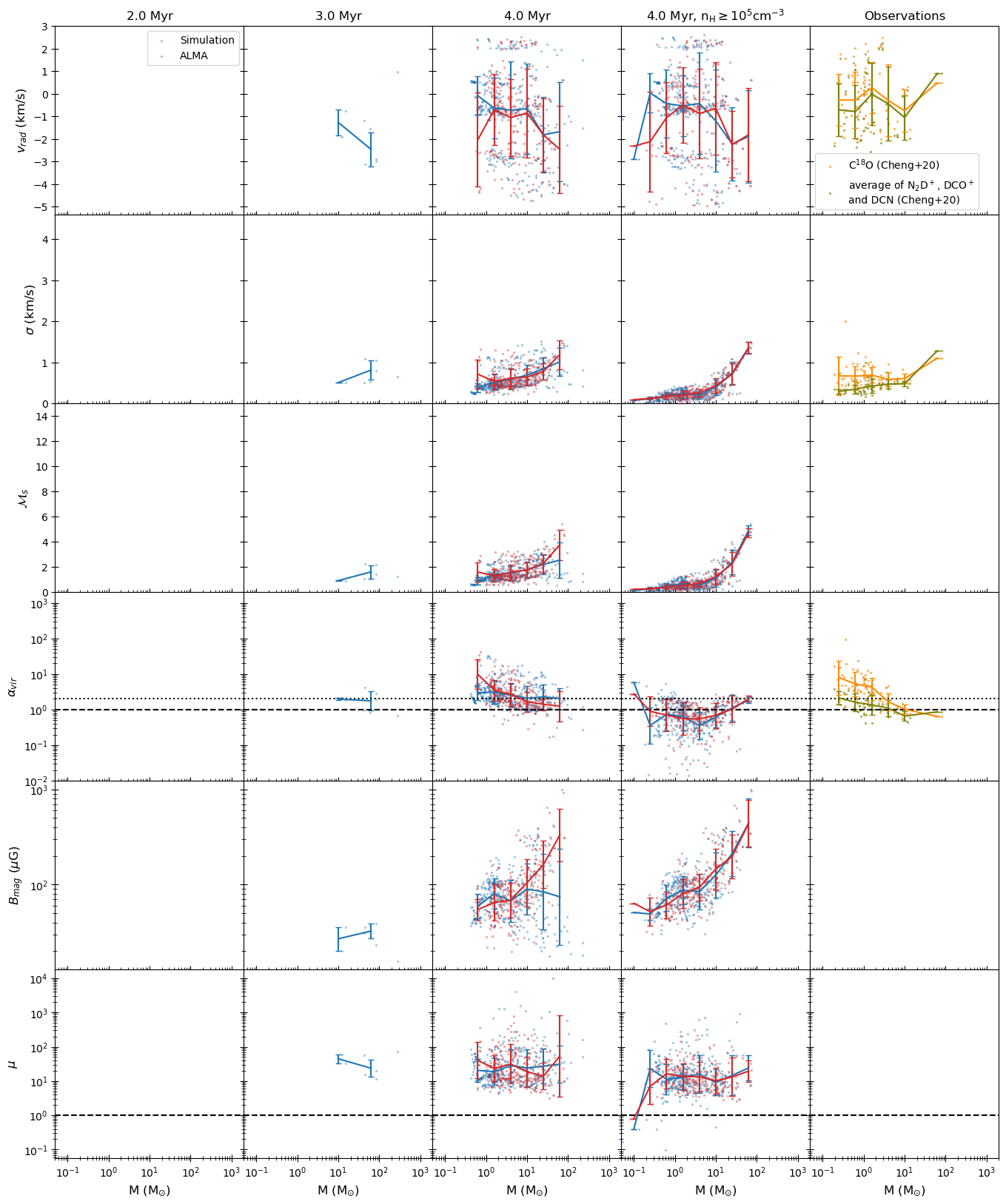}
    \caption{As Figure~\ref{fig:corekin_col}, but now for the non-colliding case.}
    \label{fig:corekin_nocol}
\end{figure*}

Figure~\ref{fig:corekin_nocol} shows the same properties as in Figure~\ref{fig:corekin_col}, but now for the non-colliding case. 
The radial velocities are distributed in a similar range as that of the colliding case, but the non-colliding case has more discrete groupings of sub-clusters. For the velocity dispersion, it also shows that more massive cores tend to have higher velocity dispersions if a density threshold is applied. Otherwise, the correlation is weak. Due to the lower velocity dispersions, the Mach numbers have a narrower range, i.e., up to $\sim$5, than the colliding case. 
Most cores are also supervirial, as in the colliding case, if there is no density threshold applied. However, applying a density threshold causes almost all cores to become subvirial. 

Comparing with the G286 data from \citet{Cheng2020}, there are two ways to estimate the velocities and the velocity dispersions: (1) measurement in $\rm C^{18}O(2-1)$, (2) the average of measurements in $\rm N_2D^+(3-2)$, $\rm DCO^+(3-2)$, and $\rm DCN(3-2)$, which is expected to be a better tracer of denser material. These two series of data are plotted in orange and olive in the fifth columns of Figures~\ref{fig:corekin_col} and \ref{fig:corekin_nocol}. To compare radial velocities, the G286 data have been subtracted by the average of the population. The two distributions of the G286 data are narrower than both the colliding and non-colliding simulation cases. The G286 population also does not show a clear trend of increasing velocity dispersion with mass. 
For the velocity dispersion, $\rm C^{18}O$ data fall in a similar range as the non-colliding case without density threshold. In contrast, the average of $\rm N_2D^+$, $\rm DCO^+$, and $\rm DCN$ is more similar to the results with density threshold, which is expected if they trace dense gas. 
For the virial parameters measured from $\rm C^{18}O$, \citet{Cheng2020} find that 5/74 are subvirial, 22/74 are gravitationally bound and 52/74 are unbound (using, for simplicity and consistency the condition $\alpha_{\rm vir}>2$). When using the average of the dense gas tracers, these fractions are: 20/55, 40/55 and 15/55, respectively. The observational result that $\sim30\%$ of the cores appear to be unbound when using dense gas tracers is very similar to the fraction found in the GMC-GMC collision simulation selecting core material with the high-density threshold. This could be interpreted as indirect evidence in support of a cloud collision scenario (or other scenario involving disturbed molecular gas kinematics) for the triggering of star formation in the G286 protocluster.



\subsection{Core Magnetic Fields}


Since magnetic fields can provide additional support to cores against collapse, we also examine the magnetic field properties in Figures~\ref{fig:corekin_col} and \ref{fig:corekin_nocol}. The fifth rows show the mass-weighted magnetic field strength along the line of sight ($z^\prime$) inside the cores. In both colliding and non-colliding cases, we see the magnetic field strength is approximately proportional to the core mass in late stages, especially for the ALMA-filtered cores. The strength can range from several tens of $\rm \mu G$ to several mG. The most massive cores in the colliding case have field strengths of around 2~mG. In the ALMA-filtered case, the values do not change much, even if the density threshold is included, showing that the dense gas already makes the dominant contribution. 

The mass-to-flux ratio provides a way to estimate the capability of magnetic fields to support the cores. The normalized mass-to-flux ratio can be defined as \citep{Mouschovias1976}: 
\begin{equation}
\mu_B = \sqrt{63 G}\frac{M}{\Phi_B},
\end{equation}
where $\displaystyle \Phi_B = \int_{S\cap (\vect{B}\cdot d\vect{A} > 0)} \vect{B}\cdot d\vect{A}$ is the magnetic flux of a core within surface $S$ and $M$ is the enclosed mass. Here $1/\sqrt{63G}$ is the critical value of the un-normalized mass-to-flux ratio. In projection, the mass-to-flux ratio can be calculated as: 
\begin{equation}
\mu_B = \sqrt{63G}\frac{M}{A_c\langle B_{\rm los}\rangle_M},
\end{equation}
where $A_c$ is the area of the core in projection and $\langle B_{\rm los}\rangle_M$ is the mass-weighted magnetic field along the line of sight. As Zeeman splitting measurements of the magnetic field strength only provide the line-of-sight component of the field \citep[e.g.,][]{Crutcher1999}, this formula provides a reasonable way to compare with observations.

In the last rows of Figures~\ref{fig:corekin_col} and \ref{fig:corekin_nocol}, we plot the mass-to-flux ratio along the $z^\prime$ axis. The values of mass-to-flux ratio often fall in a range from 10 to 100, no matter at which stage the simulation has reached and no matter whether the clouds are colliding or not. Even if we consider application of a density threshold, the values are only about a factor of two lower. Since the magnetic field flux depends on the line of sight, we also examine the mass-to-flux ratio as viewed along the $x^\prime$ and $y^\prime$ axes. However, these show similar behavior as our results along the $z^\prime$ axis. We conclude that the magnetic fields do not play an important role for supporting the cores in this simulation. However, we note that these simulations are based on the weakest initial $B-$field case of 10 $\mu$G with the GMC collision series \citep[see][]{Wu2020}. A future work will examine cores formed from GMCs that have stronger initial $B-$field strengths.




\subsection{Virial Parameter of the Protocluster}

\begin{figure*}
    \centering
    \includegraphics[width=0.9\linewidth]{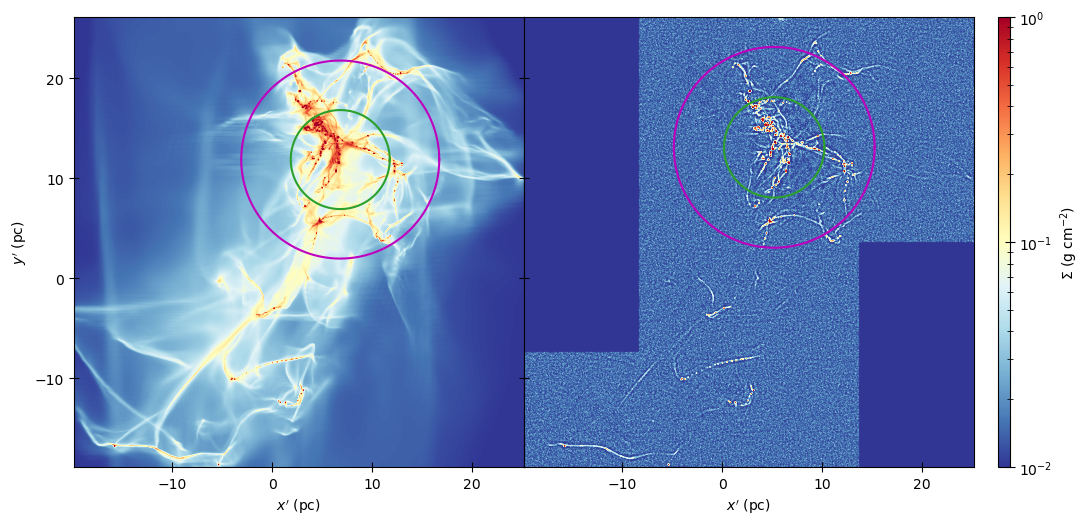}
    \caption{The selection of cluster in original simulation data (left) and ALMA synthetic observation (right). The green circle indicates the region of one $R_{\rm median}$ (see text) and purple circle indicates the region of two $R_{\rm median}$.}
    \label{fig:cluster}
\end{figure*}

In the colliding case, the two clouds have formed a large ``protocluster'' by the end of the simulation. The protocluster may have some properties reflecting the collision history. Therefore, we examine the virial parameter of the whole cluster. 
As a definition for the cluster we consider that cores are included in the cluster if they are within a distance $R_{\rm cluster}$ from the center-of mass of the cores and consider two cases, $R_{\rm cluster} = R_{\rm median}$ and $R_{\rm cluster} = 2R_{\rm median}$, where $R_{\rm median}$ is the median distance of cores from the center of mass of all the cores. In Figure~\ref{fig:cluster}, these two radii are displayed with green and purple circles, respectively. 
Since the number and location of cores change after ALMA filtering, we recompute the center of mass and the cluster radii for this case. We also consider the influence of density thresholds in this analysis.


\begin{figure*}
    \centering
    \includegraphics[width=\linewidth]{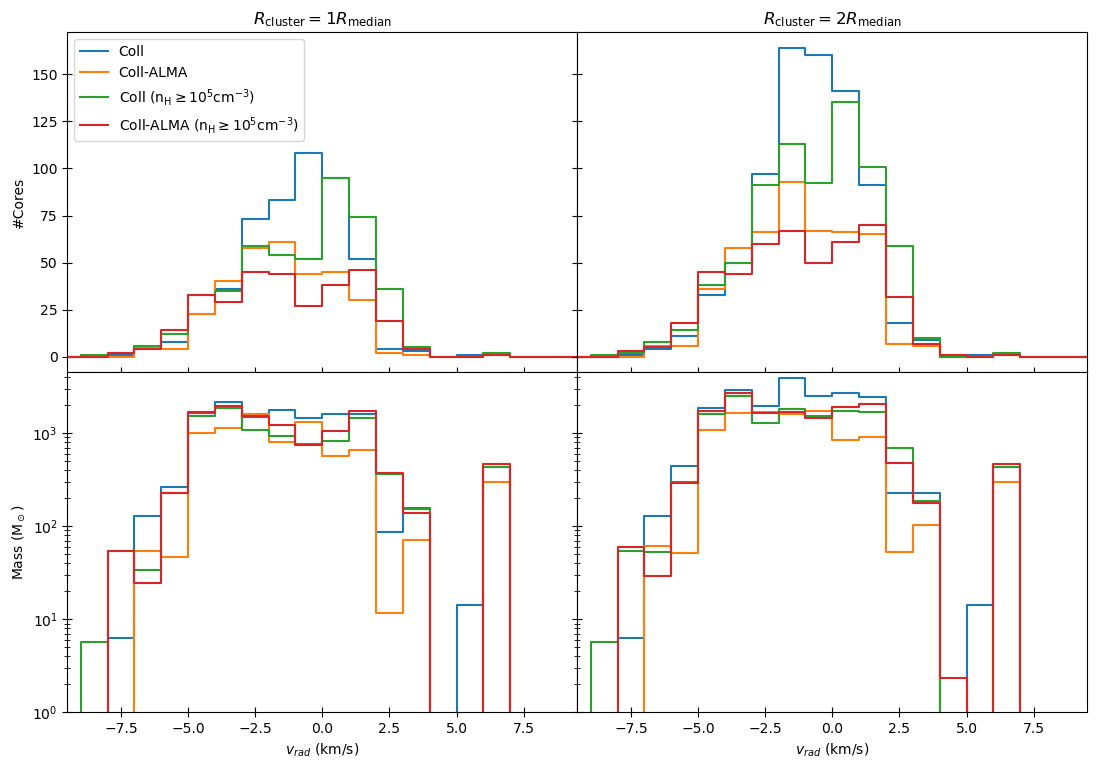}
    \caption{Histogram of radial velocity. The blue, orange, green, red lines show the cases of 4~Myr, 4~Myr (ALMA), 4~Myr ($\rm n_H \geq 10^5$), 4~Myr (ALMA, $\rm n_H \geq 10^5$), respectively. The top row shows the distribution of the number of cores and the bottom shows the distribution of mass. The left column show the result of the cores selected by 1 $R_{\rm median}$ and the right column shows the result of the cores selected by 2 $R_{\rm median}$.}
    \label{fig:clustervrad}
\end{figure*}

We start from the core velocities that have been shown in Figure~\ref{fig:vrad}. The velocity distributions of the cores selected in the clusters are plotted in Figure~\ref{fig:clustervrad}. The top row shows the distribution of cores and the bottom row shows the distribution of mass.

With the velocity of each core, we calculate the mass-weighted mean velocity of the cluster and its one-dimensional velocity dispersion. Considering within $R_{\rm median}$ in the original and ALMA-filtered data, we obtain velocity dispersions of 2.57 km/s and 2.50 km/s, respectively. Extending to $2R_{\rm median}$, we obtain 2.34 km/s and 2.33 km/s in these cases. If the density threshold $n_{\rm H} \geq 10^5 \:{\rm cm}^{-3}$ is applied, the above values change to 2.89, 2.76, 2.63 and 2.59 km/s, respectively.

The virial parameter is again estimated by Equation~\ref{eqn:virial}, where here the $R$ is the corresponding cluster radius. For the mass, we calculate the total mass enclosed in the cluster and obtain 59,842$\:M_\odot$ and 15,542$\:M_\odot$ for the original and ALMA-filtered data within one $R_{\rm median}$. We find 126,414$\:M_\odot$ and 36,785$\:M_\odot$ for these cases out to two $R_{\rm median}$. If the same density threshold is applied, the values drop to 16,719$\:M_\odot$, 17,875$\:M_\odot$, 22,973$\:M_\odot$ and 23,063$\:M_\odot$, respectively.

The results of velocity dispersion and virial parameter are listed Table~\ref{tab:clustervirial}. We see that the 1-D velocity dispersion usually becomes slightly smaller (by $<$10\%) when we adopt the larger radius. The virial parameters also tend to be smaller if there is no density threshold applied, mainly because of the significant amount of mass at lower densities.
The protocluster appears to be supervirial if measured from the ALMA-filtered image or with a density threshold applied (since then a lot of the total mass is not detected or counted). This is the case even though the cluster is actually subvirial in the original simulation. For comparison, \citet{Cheng2020} claim that the G286 protocluster is likely to be subvirial in its main substructures, while at the same time the whole cluster is close to virial equilibrium. Note, they inferred the total mass from single dish observations, so are not missing mass from interferometric filtering.

\begin{table*}
    \centering
    \caption{Velocity dispersion and virial parameter of the protocluster}
    \begin{tabular}{lcccccc}
        Model & $\sigma_1$ (km/s) & $\sigma_{\rm vir,1}$ (km/s) & $\alpha_{\rm vir,1}$ & $\sigma_2$ (km/s) & $\sigma_{\rm vir,2}$ (km/s) & $\alpha_{\rm vir,2}$\\
        \hline
        4~Myr & 2.573 & 3.224 & 0.637 & 2.344 & 3.314 & 0.505 \\
        4~Myr (ALMA) & 2.489 & 1.632 & 2.326 & 2.328 & 1.775 & 1.719 \\
        4~Myr ($n_{\rm H} \geq 10^5\:{\rm cm}^{-3}$) & 2.892 & 1.737 & 2.772 & 2.634 & 1.440 & 3.346 \\
        4~Myr (ALMA, $n_{\rm H} \geq 10^5\:{\rm cm}^{-3}$) & 2.778 & 1.758 & 2.497 & 2.606 & 1.412 & 3.407 \\
    \end{tabular}
    \bigskip
    \\
    \raggedright
    Notes. $\sigma_1$ and $\alpha_{\rm vir,1}$ are the velocity dispersion and the virial parameter found within $R_{\rm median}$ (see text). $\sigma_{\rm vir,1}$ is the required level of velocity dispersion for the protocluster to be in virial equilibrium. Similar definitions apply for $\sigma_2$, $\sigma_{\rm vir,2}$ and $\alpha_{\rm vir,2}$, which are the values calculated within $2R_{\rm median}$.
    \label{tab:clustervirial}
\end{table*}

\section{Discussion and Conclusions}
\label{sec:discussion}

We have performed an analysis of the core mass function (CMF) arising from colliding and non-colliding giant molecular clouds, with a focus on cores identified by dendrogram in projected mass surface density maps. In our fiducial case, we set a minimum mass surface density threshold of $\Sigma_{\rm min} = 0.1\,{\rm g\,cm}^{-2}$, a minimum mass surface density increment $\delta_{\rm min} = 0.025\,{\rm g\,cm}^{-2}$ and a minimum area $A_{\rm min} = 2$~pixels, equivalent to an area of $2.4\times10^{-4}\:{\rm pc}^{2}$. We find that, for the colliding case, the CMF is typically relatively flat between $1\,M_\odot$ and $10\,M_\odot$. It can be fit, approximately, by a power-law for $M \geq\:10 M_\odot$, with this index being $\alpha_{10}\simeq 0.7$ after 4~Myr, which is top-heavy compared to a Salpeter mass function that has $\alpha=1.35$.
For the non-colliding case, cores take longer to form and do so in fewer numbers.
At 4~Myr, we see that the CMF follows a moderately steeper distribution than the colliding case in all the mass ranges considered: for example, at 4~Myr is has $\alpha_{10}\simeq 0.8$. 

To understand the influence of dendrogram parameters, we also examined the CMFs found with different minimum mass surface densities, minimum mass surface density increments and minimum areas. The resulting CMFs do not change significantly at the high-mass end. However, these choices have significant influence at the low-mass end, i.e., $< 5 M_\odot$. 

Since ALMA observations tend to miss large scale structures, we applied CASA {\it simobserve} and {\it simanalyze} tasks to obtain synthetic ``ALMA-filtered'' observational results. We examined these results for the colliding case at 2, 3 and 4~Myr and for the non-colliding case at 4~Myr. 
The general effect of ALMA-filtering reduces the mass estimates of cores, so that the CMFs have a peak at a lower mass around $1\:M_\odot$. As a result, the power law indices in the range from $1\:M_\odot$ to 10 $M_\odot$ become much steeper in the colliding case and the overall CMF is better described by a single power law. However, the high-mass end index remains close to the previous value, i.e., $\alpha_{10}\simeq 0.7$.
ALMA-filtering applied to the non-colliding simulation causes a steepening of the high-end index to $\alpha_{10}\simeq 1.1$, cloer to the Salpeter value, although uncertainties are larger due to smaller numbers of massive cores.


Another factor that may influence measurement of core masses and the CMF is the presence of low-density ambient gas, which contributes to the mass surface density, but does not belong to the gravitationally bounded structure. To understand its influence, we fixed the contours of the cores identified in the original data and only counted gas along the line of sight above certain density thresholds, considering cases of $n_{\rm H} \geq 10^4$ and $10^5\:{\rm cm}^{-3}$. We examined the influence of this on the CMFs at 4~Myr. Core masses, as expected are reduced, and the CMF power law indices tend to become steeper.

We have compared our results with the CMFs from the observational studies of \citet{Cheng2018, Liu2018} and \citet{ONeill2021}, which used similar methods to identify cores. Overall, especially for the larger samples of cores in the multi-region studies of \citet{Liu2018} and \citet{ONeill2021}, we can find examples of ALMA-filtered CMFs from both colliding and non-colliding simulations that are consistent with the ``raw'' CMFs derived from these studies, which we consider to be the fairest comparison. While it is promising to find such consistency, this also means that we are not able to favor between the colliding and non-colliding scenarios. Future work that examines are broader variety of core properties and also compares to simulations that explore a wider range of parameter space (e.g., collision velocity and initial GMC magnetic field strength) will be needed for progress in this area.

Along these lines, we have also examined physical properties, other than mass, of the identified cores in our simulations. 
At the beginning of the colliding case, the core radius is proportional to the core mass, and the mass surface density is approximately constant, resulting in the estimated volume density being inversely proportional to the core mass. In contrast, by 4~Myr, massive cores have mass surface density proportional to core mass and the volume density is approximately constant. Synthetic ALMA observations modify these results further.

Comparing with observational data, the simulated cores tend to be larger and have lower densities. A potential cause of this effect is that the star-forming regions probed by the observational studies are typically closer than our adopted fiducial distance of 5~kpc and so are able to resolve smaller scales that probed in our simulations. Simulations with higher spatial resolution are needed to assess this aspect. However, another potential effect is that the observed cores are already protostellar sources, i.e., with a significant protostellar mass and associated heating that acts to concentrate the mm continuum flux that is used in the observational definition of core sizes. Future work can investigate this aspect by selecting samples of pre-stellar cores. For distant, crowded regions, one promising method for this is to use deuterated species, especially $\rm N_2D^+$ \citep[e.g.,][]{2013ApJ...779...96T,2017ApJ...834..193K}, ideally coupled with accurate (temperature-independent) estimates of mass surface density that are most readily achieved from mid-infrared extinction mapping \citep{2012ApJ...754....5B}. An alternative approach would be to implement sub-grid models of protostellar cores in the simulations that induce local heating and associated enhanced mm flux emission. However, such models involve significant uncertainties in their implementation.



Considering temperature, our simulated cores without ALMA-filtering tend to have temperatures that grow in proportion to core mass. However, ALMA filtering induces an opposite relation of temperature declining with mass. Applying a density threshold for core definition also leads to a major change, with core temperatures becoming much cooler, closer to 10~K.


From the kinematic and dynamical aspect, we find that magnetic fields in our simulations are not lending significant support to the cores. However, this may change for cases in which the initial GMCs are more strongly magnetized, which will be investigated in a future study. Most of our simulated cores are supervirial if a density threshold is not applied, no matter whether the cores are identified in the original data or after ALMA filtering. However, about half of cores in the colliding case are subvirial if the density threshold $n_{\rm H} \geq 10^5\:{\rm cm}^{-3}$ is applied. In the non-colliding case, most ($\sim70\%$) of the cores are subvirial when selected with this threshold. The fraction of unbound ($\alpha_{\rm vir}>2$) cores selected with the density threshold of $n_{\rm H} \geq 10^5\:{\rm cm}^{-3}$ is most sensitive to whether (28\%) or not (6.5\%) the cores formed from a GMC-GMC collision. A comparison against observational data for this unbound fraction in G286, which is $\sim 30\%$ \citep{Cheng2020}, is tentative evidence in favor of cloud collisions being involved in the triggering of star formation in this system.

On larger scales, the dynamical state of the protocluster of cores formed via GMC-GMC collision is intrinsically subvirial, but appears to be supervirial if the total mass measurement is affected by observations that miss mass on large scales or at low densities.



\section*{Acknowledgements}

The simulations were performed on resources provided by the Swedish National Infrastructure for Computing (SNIC) at C3SE. JCT acknowledges support from VR grant 2017-04522 (Eld ur is) and ERC Advanced Grant 788829 (MSTAR). CJH acknowledges the valuable discussion with Chi-Yan Law and Yao-Lun Yang. This research made use of astrodendro, a Python package to compute dendrograms of Astronomical data (http://www.dendrograms.org/), and yt \citep[https://yt-project.org/,][]{Turk2011} to analyse simulation data.

\section*{Data Availability}

The data underlying this article will be shared on reasonable request to the corresponding author.



\bibliographystyle{mnras}
\bibliography{main} 

\begin{thebibliography}{}
\makeatletter
\relax
\def\mn@urlcharsother{\let\do\@makeother \do\$\do\&\do\#\do\^\do\_\do\%\do\~}
\def\mn@doi{\begingroup\mn@urlcharsother \@ifnextchar [ {\mn@doi@}
  {\mn@doi@[]}}
\def\mn@doi@[#1]#2{\def\@tempa{#1}\ifx\@tempa\@empty \href
  {http://dx.doi.org/#2} {doi:#2}\else \href {http://dx.doi.org/#2} {#1}\fi
  \endgroup}
\def\mn@eprint#1#2{\mn@eprint@#1:#2::\@nil}
\def\mn@eprint@arXiv#1{\href {http://arxiv.org/abs/#1} {{\tt arXiv:#1}}}
\def\mn@eprint@dblp#1{\href {http://dblp.uni-trier.de/rec/bibtex/#1.xml}
  {dblp:#1}}
\def\mn@eprint@#1:#2:#3:#4\@nil{\def\@tempa {#1}\def\@tempb {#2}\def\@tempc
  {#3}\ifx \@tempc \@empty \let \@tempc \@tempb \let \@tempb \@tempa \fi \ifx
  \@tempb \@empty \def\@tempb {arXiv}\fi \@ifundefined
  {mn@eprint@\@tempb}{\@tempb:\@tempc}{\expandafter \expandafter \csname
  mn@eprint@\@tempb\endcsname \expandafter{\@tempc}}}

\bibitem[\protect\citeauthoryear{{Bertoldi} \& {McKee}}{{Bertoldi} \&
  {McKee}}{1992}]{1992ApJ...395..140B}
{Bertoldi} F.,  {McKee} C.~F.,  1992, \mn@doi [\apj] {10.1086/171638}, \href
  {https://ui.adsabs.harvard.edu/abs/1992ApJ...395..140B} {395, 140}

\bibitem[\protect\citeauthoryear{Bisbas et~al.,}{Bisbas
  et~al.}{2018}]{Bisbas2018}
Bisbas T.~G.,  et~al., 2018, \mn@doi [Monthly Notices of the Royal Astronomical
  Society: Letters] {10.1093/mnrasl/sly039}, 478, L54

\bibitem[\protect\citeauthoryear{Brummel-Smith et~al.,}{Brummel-Smith
  et~al.}{2019}]{Brummel-Smith2019}
Brummel-Smith C.,  et~al., 2019, \mn@doi [Journal of Open Source Software]
  {10.21105/joss.01636}, 4, 1636

\bibitem[\protect\citeauthoryear{{Butler} \& {Tan}}{{Butler} \&
  {Tan}}{2012}]{2012ApJ...754....5B}
{Butler} M.~J.,  {Tan} J.~C.,  2012, \mn@doi [\apj]
  {10.1088/0004-637X/754/1/510.48550/arXiv.1205.2391}, \href
  {https://ui.adsabs.harvard.edu/abs/2012ApJ...754....5B} {754, 5}

\bibitem[\protect\citeauthoryear{Chen \& Ostriker}{Chen \&
  Ostriker}{2018}]{Chen2018}
Chen C.~Y.,  Ostriker E.~C.,  2018, \mn@doi [arXiv] {10.3847/1538-4357/aad905},
  865, 34

\bibitem[\protect\citeauthoryear{Cheng, Tan, Liu, Kong, Lim, Andersen  \&
  Rio}{Cheng et~al.}{2018}]{Cheng2018}
Cheng Y.,  Tan J.~C.,  Liu M.,  Kong S.,  Lim W.,  Andersen M.,   Rio N.~D.,
  2018, \mn@doi [The Astrophysical Journal] {10.3847/1538-4357/aaa3f1}, 853,
  160

\bibitem[\protect\citeauthoryear{Cheng, Tan, Liu, Lim  \& Andersen}{Cheng
  et~al.}{2020}]{Cheng2020}
Cheng Y.,  Tan J.~C.,  Liu M.,  Lim W.,   Andersen M.,  2020, \mn@doi [The
  Astrophysical Journal] {10.3847/1538-4357/ab879f}, 894, 87

\bibitem[\protect\citeauthoryear{Christie, Wu  \& Tan}{Christie
  et~al.}{2017}]{Christie2017}
Christie D.,  Wu B.,   Tan J.~C.,  2017, \mn@doi [The Astrophysical Journal]
  {10.3847/1538-4357/aa8a99}, 848, 50

\bibitem[\protect\citeauthoryear{Crutcher}{Crutcher}{1999}]{Crutcher1999}
Crutcher R.~M.,  1999, \mn@doi [The Astrophysical Journal] {10.1086/307483},
  520, 706

\bibitem[\protect\citeauthoryear{Dobbs, Pringle  \& Duarte-Cabral}{Dobbs
  et~al.}{2015}]{Dobbs2015}
Dobbs C.~L.,  Pringle J.~E.,   Duarte-Cabral A.,  2015, \mn@doi [Monthly
  Notices of the Royal Astronomical Society] {10.1093/mnras/stu2319}, 446, 3608

\bibitem[\protect\citeauthoryear{{Draine}}{{Draine}}{2011}]{2011piim.book.....D}
{Draine} B.~T.,  2011, {Physics of the Interstellar and Intergalactic Medium}

\bibitem[\protect\citeauthoryear{{Fujita} et~al.,}{{Fujita}
  et~al.}{2017}]{2017arXiv171101695F}
{Fujita} S.,  et~al., 2017, arXiv e-prints, p. arXiv:1711.01695

\bibitem[\protect\citeauthoryear{Fukui et~al.,}{Fukui et~al.}{2014}]{Fukui2014}
Fukui Y.,  et~al., 2014, \mn@doi [Astrophysical Journal]
  {10.1088/0004-637X/780/1/36}, 780

\bibitem[\protect\citeauthoryear{Furukawa, Dawson, Ohama, Kawamura, Mizuno,
  Onishi  \& Fukui}{Furukawa et~al.}{2009}]{Furukawa2009}
Furukawa N.,  Dawson J.~R.,  Ohama A.,  Kawamura A.,  Mizuno N.,  Onishi T.,
  Fukui Y.,  2009, \mn@doi [Astrophysical Journal]
  {10.1088/0004-637X/696/2/L115}, 696, 115

\bibitem[\protect\citeauthoryear{{Kong}, {Tan}, {Caselli}, {Fontani}, {Liu}  \&
  {Butler}}{{Kong} et~al.}{2017}]{2017ApJ...834..193K}
{Kong} S.,  {Tan} J.~C.,  {Caselli} P.,  {Fontani} F.,  {Liu} M.,   {Butler}
  M.~J.,  2017, \mn@doi [\apj]
  {10.3847/1538-4357/834/2/19310.48550/arXiv.1609.06008}, \href
  {https://ui.adsabs.harvard.edu/abs/2017ApJ...834..193K} {834, 193}

\bibitem[\protect\citeauthoryear{Li, Tan, Christie, Bisbas  \& Wu}{Li
  et~al.}{2018}]{Li2018}
Li Q.,  Tan J.~C.,  Christie D.,  Bisbas T.~G.,   Wu B.,  2018, \mn@doi
  [Publications of the Astronomical Society of Japan] {10.1093/pasj/psx136},
  70, 1

\bibitem[\protect\citeauthoryear{Liu, Tan, Cheng  \& Kong}{Liu
  et~al.}{2018}]{Liu2018}
Liu M.,  Tan J.~C.,  Cheng Y.,   Kong S.,  2018, \mn@doi [The Astrophysical
  Journal] {10.3847/1538-4357/aacb7c}, 862, 105

\bibitem[\protect\citeauthoryear{{McMullin}, {Waters}, {Schiebel}, {Young}  \&
  {Golap}}{{McMullin} et~al.}{2007}]{2007ASPC..376..127M}
{McMullin} J.~P.,  {Waters} B.,  {Schiebel} D.,  {Young} W.,   {Golap} K.,
  2007, in {Shaw} R.~A.,  {Hill} F.,   {Bell} D.~J.,  eds,  Astronomical
  Society of the Pacific Conference Series Vol. 376, Astronomical Data Analysis
  Software and Systems XVI. p.~127

\bibitem[\protect\citeauthoryear{Mouschovias \& {Spitzer, L.}}{Mouschovias \&
  {Spitzer, L.}}{1976}]{Mouschovias1976}
Mouschovias T.~C.,  {Spitzer, L.} J.,  1976, Astrophysical Journal, 210, 326,
  327

\bibitem[\protect\citeauthoryear{O'Neill, Cosentino, Tan, Cheng  \&
  Liu}{O'Neill et~al.}{2021}]{ONeill2021}
O'Neill T.~J.,  Cosentino G.,  Tan J.~C.,  Cheng Y.,   Liu M.,  2021, \mn@doi
  [The Astrophysical Journal] {10.3847/1538-4357/ac062d}, 916, 45

\bibitem[\protect\citeauthoryear{Offner, Clark, Hennebelle, Bastian, Bate,
  Hopkins, Moreaux  \& Whitworth}{Offner et~al.}{2014}]{Offner2014}
Offner S. S.~R.,  Clark P.~C.,  Hennebelle P.,  Bastian N.,  Bate M.~R.,
  Hopkins P.~F.,  Moreaux E.,   Whitworth A.~P.,  2014, \mn@doi [Protostars and
  Planets VI] {10.2458/azu_uapress_9780816531240-ch003}

\bibitem[\protect\citeauthoryear{{Ossenkopf} \& {Henning}}{{Ossenkopf} \&
  {Henning}}{1994}]{1994A&A...291..943O}
{Ossenkopf} V.,  {Henning} T.,  1994, \aap, \href
  {https://ui.adsabs.harvard.edu/abs/1994A&A...291..943O} {291, 943}

\bibitem[\protect\citeauthoryear{Rosolowsky, Pineda, Kauffmann  \&
  Goodman}{Rosolowsky et~al.}{2008}]{Rosolowsky2008}
Rosolowsky E.~W.,  Pineda J.~E.,  Kauffmann J.,   Goodman A.~A.,  2008, \mn@doi
  [The Astrophysical Journal] {10.1086/587685}, 679, 1338

\bibitem[\protect\citeauthoryear{{Salpeter}}{{Salpeter}}{1955}]{1955ApJ...121..161S}
{Salpeter} E.~E.,  1955, \mn@doi [\apj] {10.1086/145971}, \href
  {https://ui.adsabs.harvard.edu/abs/1955ApJ...121..161S} {121, 161}

\bibitem[\protect\citeauthoryear{Scoville, Sanders  \& Clemens}{Scoville
  et~al.}{1986}]{Scoville1986}
Scoville N.~Z.,  Sanders D.~B.,   Clemens D.~P.,  1986, \mn@doi [The
  Astrophysical Journal] {10.1086/184785}, 310, L77

\bibitem[\protect\citeauthoryear{Suwannajak, Tan  \& Leroy}{Suwannajak
  et~al.}{2014}]{Suwannajak2014}
Suwannajak C.,  Tan J.~C.,   Leroy A.~K.,  2014, \mn@doi [Astrophysical
  Journal] {10.1088/0004-637X/787/1/68}, 787

\bibitem[\protect\citeauthoryear{Tan}{Tan}{2000}]{Tan2000}
Tan J.~C.,  2000, \mn@doi [The Astrophysical Journal] {10.1086/308905}, 536,
  173

\bibitem[\protect\citeauthoryear{Tan}{Tan}{2010}]{Tan2010}
Tan J.~C.,  2010, \mn@doi [Astrophysical Journal Letters]
  {10.1088/2041-8205/710/1/L88}, 710, 88

\bibitem[\protect\citeauthoryear{{Tan}, {Kong}, {Butler}, {Caselli}  \&
  {Fontani}}{{Tan} et~al.}{2013}]{2013ApJ...779...96T}
{Tan} J.~C.,  {Kong} S.,  {Butler} M.~J.,  {Caselli} P.,   {Fontani} F.,  2013,
  \mn@doi [\apj] {10.1088/0004-637X/779/2/9610.48550/arXiv.1303.4343}, \href
  {https://ui.adsabs.harvard.edu/abs/2013ApJ...779...96T} {779, 96}

\bibitem[\protect\citeauthoryear{Tan, Beltran, Caselli, Fontani, Fuente,
  Krumholz, McKee  \& Stolte}{Tan et~al.}{2014}]{Tan2014}
Tan J.~C.,  Beltran M.~T.,  Caselli P.,  Fontani F.,  Fuente A.,  Krumholz
  M.~R.,  McKee C.~F.,   Stolte A.,  2014, \mn@doi [International Astronomical
  Union Colloquium] {10.2458/azu_uapress_9780816531240-ch007}, 140, 176

\bibitem[\protect\citeauthoryear{Tasker \& Tan}{Tasker \&
  Tan}{2009}]{Tasker2009}
Tasker E.~J.,  Tan J.~C.,  2009, \mn@doi [Astrophysical Journal]
  {10.1088/0004-637X/700/1/358}, 700, 358

\bibitem[\protect\citeauthoryear{Turk, Smith, Oishi, Skory, Skillman, Abel  \&
  Norman}{Turk et~al.}{2011}]{Turk2011}
Turk M.~J.,  Smith B.~D.,  Oishi J.~S.,  Skory S.,  Skillman S.~W.,  Abel T.,
  Norman M.~L.,  2011, \mn@doi [Astrophysical Journal, Supplement Series]
  {10.1088/0067-0049/192/1/9}, 192

\bibitem[\protect\citeauthoryear{Wang \& Abel}{Wang \& Abel}{2009}]{Wang2009}
Wang P.,  Abel T.,  2009, \mn@doi [The Astrophysical Journal]
  {10.1088/0004-637X/696/1/96}, 696, 96

\bibitem[\protect\citeauthoryear{Wang, Abel  \& Zhang}{Wang
  et~al.}{2008}]{Wang2008}
Wang P.,  Abel T.,   Zhang W.,  2008, \mn@doi [The Astrophysical Journal
  Supplement Series] {10.1086/529434}, 176, 467

\bibitem[\protect\citeauthoryear{Wu, Tan, Nakamura, Loo, Christie  \&
  Collins}{Wu et~al.}{2017a}]{Wu2017}
Wu B.,  Tan J.~C.,  Nakamura F.,  Loo S.~V.,  Christie D.,   Collins D.,
  2017a, \mn@doi [The Astrophysical Journal] {10.3847/1538-4357/835/2/137},
  835, 137

\bibitem[\protect\citeauthoryear{Wu, Tan, Christie, Nakamura, {Van Loo}  \&
  Collins}{Wu et~al.}{2017b}]{Wu2017a}
Wu B.,  Tan J.~C.,  Christie D.,  Nakamura F.,  {Van Loo} S.,   Collins D.,
  2017b, \mn@doi [The Astrophysical Journal] {10.3847/1538-4357/aa6ffa}, 841,
  88

\bibitem[\protect\citeauthoryear{Wu, Tan, Christie  \& Nakamura}{Wu
  et~al.}{2020}]{Wu2020}
Wu B.,  Tan J.~C.,  Christie D.,   Nakamura F.,  2020, \mn@doi [The
  Astrophysical Journal] {10.3847/1538-4357/ab77b5}, 891, 168

\bibitem[\protect\citeauthoryear{Zhang \& Tan}{Zhang \& Tan}{2015}]{Zhang2015}
Zhang Y.,  Tan J.~C.,  2015, \mn@doi [Astrophysical Journal Letters]
  {10.1088/2041-8205/802/2/L15}, 802, L15

\makeatother
\end{thebibliography}








\bsp	
\label{lastpage}
\end{document}